\documentclass[journal]{IEEEtran}

\usepackage{cite}

\usepackage{amsmath}
\usepackage{mathrsfs}

\usepackage{algorithmic}

\usepackage{array}

\usepackage{booktabs}

\ifCLASSOPTIONcompsoc
\usepackage[caption=false,font=normalsize,labelfont=sf,textfont=sf]{subfig}
\else
\usepackage[caption=false,font=footnotesize]{subfig}
\fi

\usepackage{url}
\usepackage{graphicx,amsmath,amssymb,amsfonts}
\usepackage{algorithmic,algorithm}

\usepackage{setspace}

\usepackage{xcolor}

\newtheorem{remark}{\textbf{Remark}}
\newtheorem{theorem}{\textbf{Theorem}}

\newtheorem{lemma}{\textbf{Lemma}}

\newtheorem{corollary}{\textbf{Corollary}}

\newtheorem{proposition}{\textbf{Proposition}}

\renewcommand{\IEEEQED}{\IEEEQEDopen}

\makeatletter

\newcommand{\Rmnum}[1]{\expandafter\@slowromancap\romannumeral #1@}
\makeatother

\usepackage{framed} 

\usepackage{cancel}

\begin{document}
\bstctlcite{ref:BSTcontrol}

\title{Research on Resource Allocation for Efficient Federated Learning}

\author{
		Jianyang~Ren,~
		Wanli~Ni,~
		Gaofeng~Nie,~
		Hui~Tian
		\thanks{A preliminary version of this paper has been presented at the IEEE ICC Workshop on Edge Learning for 5G Mobile Networks and Beyond, virtual, June 2021\cite{ren2021joint}.
			
		J. Ren, W. Ni, G. Nie and H. Tian are with the State Key Laboratory of Networking and Switching Technology, Beijing University of Posts and Telecommunications,
		Beijing, China.}
		}
	
\maketitle

\begin{abstract}
	As a promising solution to achieve efficient learning among isolated data owners and solve data privacy issues, federated learning is receiving wide attention.
	Using the edge server as an intermediary can effectively collect sensor data, perform local model training, and upload model parameters for global aggregation.
	So this paper proposes a new framework for resource allocation in a hierarchical network supported by edge computing.
	In this framework, we minimize the weighted sum of system cost and learning cost by optimizing bandwidth, computing frequency, power allocation and subcarrier assignment.
	To solve this challenging mixed-integer non-linear problem, we first decouple the bandwidth optimization problem(P1) from the whole problem and obtain a closed-form solution.
	The remaining computational frequency, power, and subcarrier joint optimization problem(P2) can be further decomposed into two sub-problems: latency and computational frequency optimization problem(P3) and transmission power and subcarrier optimization problem(P4).
	P3 is a convex optimization problem that is easy to solve.
	In the joint optimization problem(P4), the optimal power under each subcarrier selection can be obtained first through the successive convex approximation(SCA) algorithm.
	Substituting the optimal power value obtained back to P4, the subproblem can be regarded as an assignment problem, so the Hungarian algorithm can be effectively used to solve it.
	The solution of problem P2 is accomplished by solving P3 and P4 iteratively.
	To verify the performance of the algorithm, we compare the proposed algorithm with five algorithms; namely Equal bandwidth allocation, Learning cost guaranteed, Greedy subcarrier allocation, System cost guaranteed and Time-biased algorithm.
	Numerical results show the significant performance gain and the robustness of the proposed algorithm in the face of parameter changes.
\end{abstract}

\begin{IEEEkeywords}
	 federated learning, edge computing, resource allocation
\end{IEEEkeywords}

\section{Introduction}
The past few years have seen the evolution and advancement of the Internet of Things (IoT), which enables numerous resource-limited devices to be connected with each other via the Internet\cite{atzori2010internet}.
With the development of the Internet of Things, the number of sensors has exploded and generated huge amounts of data\cite{zhang2018data}.
However, because of the embedded low processor and storage capacity, most IoT devices have limited resources for computing, storage, and communication\cite{liu2020toward}.
In order to alleviate the resource constraints of IoT devices, cloud computing technologies have been proposed to provide task offloading.
While network latency and heavy traffic burden will be incurred by offloading IoT tasks to remote cloud\cite{ansari2018mobile,mouradian2017comprehensive}.
Fortunately, Mobile Edge Computing (MEC) solves these problems by deploying servers within the radio access network (RAN) , and becomes a key technology to support latency-critical mobile and IoT applications\cite{wang2019edge}.

In addition, the development of mobile Internet technology has greatly promoted artificial intelligence so that human society has entered the intelligence era\cite{ye2020edgefed}. 
Machine Learning (ML) helps process the huge amounts of data created by IoT devices and gives IoT time to think so that it can use its technologies and ideas to deliver the desired results\cite{majumdar2019survey}.
However, traditional ML framework mainly focus on centralized data processing, which poses great challenges to data privacy protection and network throughput.
To solve these problems, a decentralized ML approach called federated learning (FL) is widely regarded as an attractive approach\cite{konevcny2016federated}.
It allows participants to use local data sets for local training, and then upload model parameters for parameter aggregation to avoid the problems of privacy leakage and network transmission pressure caused by original data transmission.
Unfortunately, FL also has its own challenges as follows.
\begin{itemize}
	\item [1)]
	The performance of federated learning (e.g., the time spent on a single round of federated learning) is affected by the allocation of resources (e.g., bandwidth, computing resources, power, etc.) in the network.
	\item [2)]
	A collaborative FL model for different participants must take into account the validity of the learning model parameters received from them\cite{khan2020resource}. 
	Therefore, the transmission error in the process of model transmission is also an important factor affecting FL performance.
\end{itemize}

Based on the important role that MEC and FL can play in the future IoT environment mentioned above, it is very promising to consider the resource optimization problem for efficient federated learning in the network environment supported by edge computing.
In fact, introducing edge servers into the network architecture is very beneficial for the IoT FL scenario.
Its main advantages can be summarized as the following.
\begin{itemize}
	\item [1)]
	In the IoT scenario, wireless sensors with extremely limited computing and storage resources are the main source of data. 
	The local processing of data by sensors will cause long latency and slow down the learning process seriously. 
	While newly introduced edge servers can collect sensor data and quickly complete local model training and data storage using powerful computing and storage capabilities.
	\item [2)]
	Wireless sensors are powered by their own batteries, so their energy is limited.
	Migrating the local model training process to the edge server helps reduce the energy consumption of the sensors, thereby improving their lifetime.
	\item [3)]
	Small base stations (SBS) co-deployed with edge servers have more powerful communication capabilities. 
	They have a higher power budget and therefore can transfer parameters with higher power, thus mitigating the impact of transmission errors on FL learning performance.
\end{itemize}
\subsection{Related Works}
\emph{1)Resource optimization in edge computing networks:}
Resource allocation algorithms for reducing service delay and system energy consumption in edge computing networks have been studied extensively in the past few years\cite{sardellitti2015joint,zhang2016energy,wei2018greedy,yu2018energy,dab2019joint,zhao2020energy,yuhong2018energy,mao2017energy}.
In the optimization problems formulated in \cite{sardellitti2015joint} and \cite{zhang2016energy}, only the energy consumption of the system is considered.
Sardellitti \textit{et. al} \cite{sardellitti2015joint} studied the problem of energy consumption optimization in the multi-cell multi-user edge computing system.
The optimization problem considers minimizing the weighted sum of energy consumption of the whole system by optimizing the offloading sequence of tasks, the time of data transmission, the time of data transmission and the time of task execution.
A low complexity algorithm is designed based on Johnson algorithm and convex optimization technique.
In the case of single cell and multiple users, the authors in \cite{zhang2016energy} presented an optimization scheme of offloading strategy and resource allocation scheme.
While \cite{wei2018greedy,yu2018energy,dab2019joint,zhao2020energy,yuhong2018energy,mao2017energy} took both  delay and energy consumption into account.
The authors of references \cite{wei2018greedy,yu2018energy,dab2019joint,zhao2020energy} all considered the optimization of energy consumption under the constraints of the delay.
Among them, Wei \textit{et. al} \cite{wei2018greedy} aimed at maximizing energy saving, while \cite{yu2018energy,dab2019joint,zhao2020energy} aimed at minimizing the sum of energy consumption.
In addition, Zhao \textit{et. al} \cite{zhao2020energy} focused on a more complex system with heterogeneous clouds, including both edge clouds and remote clouds.
In \cite{yuhong2018energy} and \cite{mao2017energy}, two system indexes, energy consumption and delay, were directly considered in the optimization objective. 
Yu \textit{et. al} \cite{yuhong2018energy} focused on the resource optimization problem under edge cloud network architecture, while Wang \textit{et. al} \cite{mao2017energy} considered a new network framework combining wireless charging network and edge computing network.

In order to further improve the performance of the traditional edge computing architecture and increase the flexibility of the network, multi-layer mobile edge computing network was considered in \cite{wang2019joint,wang2020distributed}.
The network architecture further leverages the computing power of the layers of devices between the edge server and the remote cloud center to provide better offloading services.
The resource optimization schemes with delay and network energy consumption as optimization objectives were respectively put forward in \cite{wang2019joint} and \cite{wang2020distributed}.

\emph{2)Research on Federated Learning:}
As a new technology, federated learning is becoming more and more widely studied\cite{khan2020resource,wang2019adaptive,shi2020joint,mo2020energy,yang2020energy,jin2020design,wang2020local,chen2020fedcluster,liu2020client,luo2020hfel}.
To minimize the loss function under a given resource budget, Wang \textit{et. al} \cite{wang2019adaptive} proposed a control algorithm aiming at finding the best tradeoff between local update and global parameter aggregation.
With the same optimization objective, a joint bandwidth allocation and scheduling problem was formulated in \cite{shi2020joint}.
The joint optimization problem was then skillfully decomposed into two sub-problems: bandwidth allocation and device scheduling, and was solved effectively.
Further adding the delay of federated learning into the optimization objective,
a novel system model, namely, DFL was devised in \cite{khan2020resource} for the cognitive Internet of things (C-IoT).
In this network scenario, an integer linear programming problem was constructed by the authors to minimize the global FL cost.
Based on problem decomposition and variable relaxation, the optimization problem was transformed into two low-complexity convex optimization problems and solved by iterative method.
To realize green and efficient federated learning, the authors in \cite{mo2020energy} and \cite{yang2020energy} conducted relevant studies with the goal of optimizing system energy consumption.
Mo \textit{et. al} \cite{mo2020energy} minimized the total energy consumption of all edge devices by jointly determining transmission power, rate and CPU frequency of local computation under two network transmission protocols, NOMA and TDMA, respectively.
Considering the joint optimization of time allocation, bandwidth allocation, power, calculation frequency and learning accuracy, Yang \textit{et. al} \cite{yang2020energy} formulated the total energy consumption minimization problem with delay constraints.
In order to solve this problem, a new time minimization problem was constructed to search for the initial feasible solution and then an iterative algorithm was applied to complete the solution.
Energy consumption and learning performance of the system were jointly considered in \cite{jin2020design}. 
Jin \textit{et. al} \cite{jin2020design} respectively studied the problem of energy consumption minimization under the constraint of learning performance and the learning performance optimization problem with given energy consumption budget.

For the purpose of accelerating model training and reducing communication overhead, the authors in \cite{wang2020local,chen2020fedcluster,liu2020client,luo2020hfel} all designed hierarchical federated learning frameworks.
In \cite{wang2020local,chen2020fedcluster,liu2020client}, only the performance of the proposed architecture and the convergence of the corresponding algorithm were concerned.
While the importance of time delay and energy consumption performance was further considered in \cite{luo2020hfel}.
Specifically, Luo \textit{et. al}\cite{luo2020hfel} proposed a communication resource optimization scheme minimizing energy consumption and delay within one global iteration under the proposed hierarchical federated learning framework(HFEL).

\subsection{Motivations and Contributions}
Although researches \cite{sardellitti2015joint,zhang2016energy,wei2018greedy,yu2018energy,dab2019joint,zhao2020energy,yuhong2018energy,mao2017energy,wang2019joint,wang2020distributed} have studied resource optimization schemes in various edge computing scenarios, federated learning is not considered in all of them.

Therefore the proposed resource optimization schemes can not be directly applied to federated learning scenarios.
The studies in \cite{khan2020resource,wang2019adaptive,shi2020joint,mo2020energy,yang2020energy,jin2020design} considered resource optimization for federated learning, but they only considered one of the learning performance and energy consumption as optimization objectives, besides neither of them considered hierarchical network architecture.
For those studies that considered hierarchical network architectures\cite{wang2020local,chen2020fedcluster,liu2020client,luo2020hfel} , most of them focused on the design of federated learning architectures.
In addition, they all assumed that the terminal device has strong computing capacity, which might be imapplicable to resource-constrained wireless sensors in the IoT.

Therefore, in view of the limitations of the above research, this paper considers the federated learning scenario in the IoT supported by edge computing.
Besides, federated learning delay, training loss and system energy consumption are considered in the optimization objective of this paper.
The main contributions of this paper are summarized as follows:
\begin{itemize}
	\item [1)]
	We propose a new framework for federated learning scenarios in a hierarchical network architecture supported by edge computing. 
	This architecture takes full advantage of the computing, storage, and communication capabilities of the edge server and can be applied well in the federated learning scenario of the Internet of Things.
	\item [2)]
	We formulate a mixed integer nonlinear programming problem (MINLP) to minimize the weighted sum of system cost and learning cost.
	In order to solve this NP-hard problem, the original problem is finally decomposed into three sub-problems: bandwidth optimization (P1), computational frequency optimization (P3) and joint optimization of power and subcarriers (P4).
	P1 and P2 are convex optimization problems, which can be solved by convex optimization method.
	In P3, we can solve the optimal power under each subcarrier scheme by using SCA approximation method first, and then solve the assignment problem for determining the optimal subcarrier allocation scheme by Hungarian algorithm.
	\item [3)]
	We verify the performance of the proposed algorithm by numerical simulation.
	By comparing the performance of the proposed algorithm with the five comparison schemes, it can be seen that our proposed algorithm can bring the maximum performance improvement through the joint optimization of multiple resources.
\end{itemize}

The rest of this paper is organized as follows.
First, the system model of the network and the formulated problem is given in Section \Rmnum{2}. 
Then, the joint resource optimization problem is analyzed and solved in Section \Rmnum{3}. 
Finally, numerical results are shown in Section \Rmnum{4}, which is followed by conclusion and future works in Section \Rmnum{5}.

\section{System Model and Problem Formulation}
	\label{system}
\begin{figure} [t!]
	\centering
	\includegraphics[width=3.2 in]{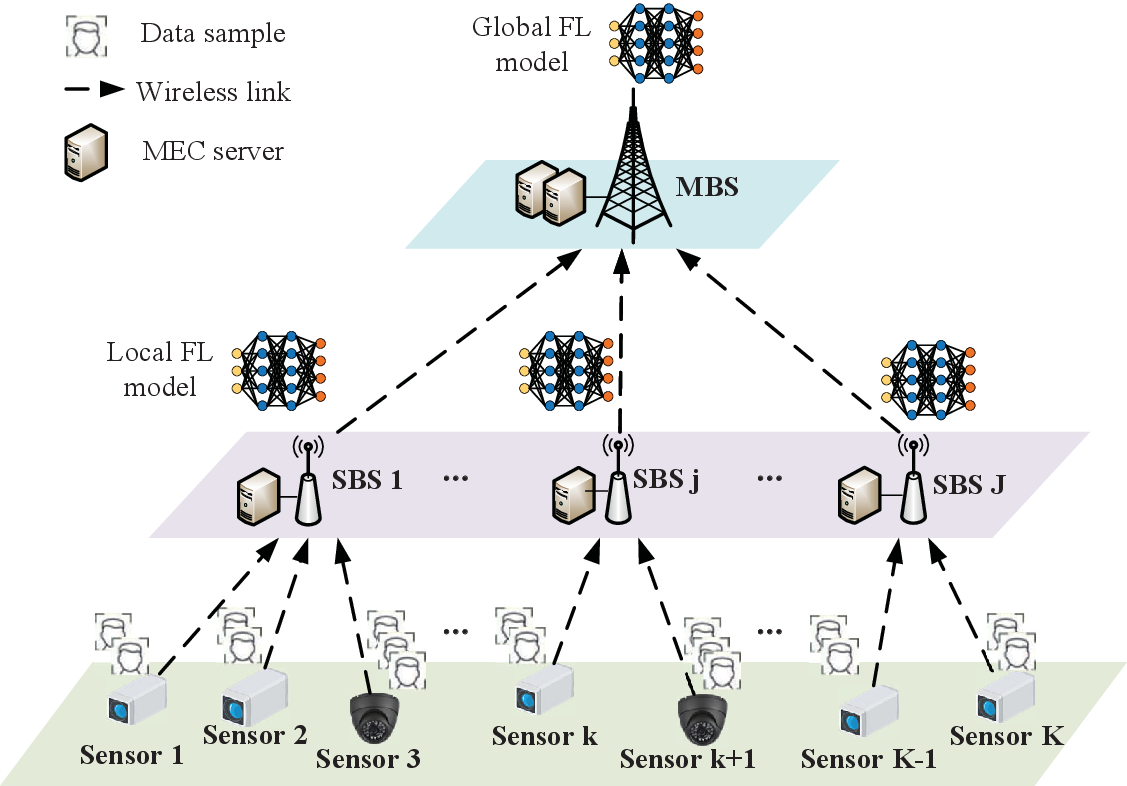}
	\caption{System model of MEC supported federated learning.}
	\label{system_model}
\end{figure}

As illustrated in Fig. \ref{system_model}, we consider a network that consists of one macro base station (MBS), $J$ small base stations (SBS) co-located with MEC servers and $K$ wireless sensors.
The sets of SBSs and sensors are denoted by $\mathcal{J} = \{ 1,2, \ldots, J\}$ and $\mathcal{K} = \{ 1,2, \ldots, K\}$, respectively. 
The set of senseors served by SBS $j$ is indexed by $\mathcal{S}_{j}$ that satisfies $\mathcal{S}_{i}\cap \mathcal{S}_{j}=\varnothing, (i\neq j)$ and $\cup_j \mathcal{S}_{j}=\mathcal{K}$.
For SBSs communicating with MBS, the total bandwidth of he MBS $B^{\rm m}$ is divided equally into $J$ orthogonal subcarriers.
To avoid interference, each subcarrier is selected for use by only one SBS.
Besides, all channels considered in this paper are regarded as Rayleigh fading channels.

Each SBS can be considered to be deployed within an enterprise or organization, so the sensors offered by the same SBS does not have the problem of privacy leakage when transmitting data through the wireless link.
In the process of completing a round of model aggregation, each SBS first collects data samples from sensors and then starts local model training. 
After the training is completed, the model is sent to MBS for model aggregation with the models of other SBSs.

\subsection{Latency Model}
Considering the access mode of frequency domain multiple access (FDMA), the achievable transmission rate of sensor $k$ served by SBS $j$ can be formulated by
\begin{equation}\label{rate_appendix_a_0}
	{R_{k}} =B_{k}^{\rm s}{\rm log}_{2}\left(1+\frac{p_k^{\rm s}h_{k}^{\rm s}}{B_k^{\rm s}N_{0}}\right),
\end{equation}
where $B_{k}^{\rm s}$ is the bandwidth allocated to sensor $k$ by SBS $j$,
$p_{k}^{\rm s}$ is the transmition power, $h_{k}^{\rm s}$ is the channel gain from sensor $k$ to SBS located in organization $j$, and
$N_{0}$ is the noise power spectrum density.
Assume that the maximum transmission power denoted by $p^{\rm max}$ is same for all sensors and the magnitude of the transmitted power $p_{k}^{\rm s}$ is proportional to the bandwidth allocated to it.
Thus, the expression for $R_k$ in (\ref{rate_appendix_a_0}) can be reformulated as: 
\begin{equation}\label{rate_appendix_a}
	{R_{k}} =B_{k}^{\rm s}{\rm log}_{2}\left(1+\frac{p^{\rm max}h_{k}^{\rm s}}{B_jN_{0}}\right).
\end{equation} 
Due to the limited bandwidth of SBS, we have: $\sum_{k\in{\mathcal{S}_{j}}}B_{k}^{\rm s}\le B_j$,where $B_j$ is the total bandwidth of SBS $j$. 
Let $t_{k}$ denote the communication time for sensor $k$ to transmit its local data set $\mathcal{D}_{k}^{\rm s}$ including $N_{k}^{\rm s}$ samples to SBS $j$.
And $D_{k}^{\rm s}$ is the size of the amount of data in $\mathcal{D}_{k}^{\rm s}$.
Thus $t_{k}$ can be characteried by
\begin{equation}\label{time_appendix_a}
	{t_{k}} =\frac{D_{k}^{\rm s}}{R_{k}}.
\end{equation}

A synchronous federated learning scenario, where all the sensors start transmitting data at the same time is cnsidered in this paper.
Therefore, the time taken by SBS $j$ receiving all the data completely is
\begin{equation}\label{time_receive_appendix_a}
	{t{_j^{\rm r}}} =\max_{k\in{\mathcal{S}_{j}}}\left(t_{k}\right).
\end{equation}

The total data collected by SBS $j$ denoted by $D_j$ is the sum of the data uploaded by all sensors within its range, which can be expressed as
\begin{equation}
	D_j =\sum_{k\in{\mathcal{S}_{j}}}D_{k}^{\rm s}.
\end{equation}
After receiving all the data, the MEC server begins training the local model immediately.
Let the positive constant $\epsilon$ (cycles/bit) represent the number of CPU cycles required for computing one bit data.
Based on the above instructions, the time $t{_j^{\rm cmp}}$ consumed by MEC server $j$ in the process of training model can be denoted by
\begin{equation}
	{t{_j^{\rm cmp}}} =\frac{\epsilon{D_j}}{f_j},
\end{equation}
where $f_j$ is the CPU frequency used by MEC server $j$ to train its local model.

After local training is completed in each SBS, SBS will send the learned model parameters $\mathcal{W}_{j}$ to MBS for aggregation with the parameters of other organizations.
In fact, in the federated learning scenario, each participant trains the same type of the model.
The model parameters that need to be transmitted from them are of similar size.
Therefore we uniformly use $D$ to represent the parameter size in this paper.
As mentioned above, the total bandwidth of the MBS $B^{\rm m}$ is divided equally into $J$ orthogonal subcarriers, so the bandwidth of each subcarrier is $B={B^{\rm m}/J}$.
Further let $C_{j,n}\in{\{0,1}\}$ denote the carrier allocation state.
Specifically, if the subcaarrier $n$ is allocated to SBS $j$, $C_{j,n}=1$, otherwise $C_{j,n}=0$.
So the condition that only one subcarrier is assigned to each SBS, and each subcarrier is assigned to only one SBS can be expressed as the following constraints:
\begin{equation}
	\sum_{j=1}^J{C_{j,n}}=1;\qquad \forall{n},
\end{equation}
\begin{equation}\label{subcarrier_constrant_2}
	\sum_{n=1}^J{C_{j,n}}=1;\qquad \forall{j}.
\end{equation}

Based on the above description, the data rate of the organization $j$ for uploading model parameters can be expressed as
\begin{equation}\label{rate_up}
	R{_{j}^{\rm up}}=\sum_{n=1}^JC{_{j,n}}B{\rm log}_{2}\left(1+\frac{p_jh_{j,n}}{B{N_{0}}}\right),
\end{equation}
where $p_j$ is the uplink transmission power of SBS $j$, since each SBS only transmits on the assigned subcarrier and the uplink channel gain of SBS $j$ on the subcarrier $n$ is denoted as $h_{j,n}$.
$R{_{j}^{\rm up}}$ in (\ref{rate_up}) is expressed as a summation of a number of terms.
However, according to constraint (\ref{subcarrier_constrant_2}), we can learn that each SBS can only select one subcarrier, that is only one of $C_{j,n}\  (\forall{n})$ is equal to one.
Therefore, the value of $R{_{j}^{\rm up}}$ is always equal to the transmission rate of SBS $j$ on the selected subcarrier.

Let $t{_{j}^{\rm up}}$ denote parameter transmission time.
It can be represented as
\begin{equation}
	t{_{j}^{\rm up}}=\frac{D}{R{_{j}^{\rm up}}}.
\end{equation}
Therefore the time delay $t{_{j}^{\rm total}}$ experienced by SBS $j$ in completing one learning process from data collection to parameter uploading are as follows:
\begin{equation}
	t{_{j}^{\rm total}}=t{_{j}^{\rm r}}+t{_{j}^{\rm cmp}}+t{_{j}^{\rm up}}.
\end{equation}

As soon as MBS receives the parameters uploaded by all SBSs, it aggregates the available parameters whose definitions will be given later in the federated learning model section.
After completing the aggregation, MBS then broadcasts the aggregated parameters to all SBSs.
In the actual scenario, due to the fast matrix operation and sufficient downlink broadcast channel bandwidth, both latency of model aggregation and downlink parameter broadcast is very small.
Therefore both of them are ignored in the study of this paper.

Since the synchronous federated learning faces straggler's dilemma issue, the time taken to perform one round of model aggregation can be formulated as
\begin{equation}
	t^{\rm one}=\max_{j\in\mathcal{J}}\left({t{_{j}^{\rm total}}}\right).
\end{equation}

\subsection{Energy Consumption Model}
In this paper, the energy consumption considered mainly occurs in three stages: sensor running, local model training in edge server and SBS model parameter transmission.
In the actual IoT scenario, the main energy consumption of the sensor is the energy consumption in the monitoring process and data transmission process.
The energy consumption in the monitoring process is mainly related to the type and working mode of the sensor, which is not considered in this paper.
While the energy of sensor $k$ consumed by the data transmission can be expressed as
\begin{equation}\label{sensor_energy}
	{e_{k}} =p_k^{\rm s}t_{k}.
\end{equation}
After substituting the expression of the relevant variable, the formula (\ref{sensor_energy}) can be converted to
\begin{equation}
	{e_{k}} =\frac{p^{\rm max}D_{k}^{\rm s}}{B{_j}{\rm log}_{2}\left(1+\frac{p^{\rm max}h_{k}^{\rm s}}{B{_j}N_{0}}\right)}.
\end{equation}
It can be seen from the above formula that when the transmission power used per unit bandwidth is constant, the transmission energy consumption of the sensor is only related to the channel state at the current moment.
Therefore, resource optimization is not beneficial to save energy consumption in the process of sensor data uploading in this case.

Next, the energy consumption of edge server $j$ in the process of local model training can be expressed as:
\begin{equation}
	{e{_j^{\rm cmp}}} =\kappa\epsilon{D{_j}}\left(f{_j}\right)^{2},
\end{equation}
where $\kappa$ is effective switched capacitance.
It is a positive constant that only depends on the structure of the chip.

Finally, the energy consumption of each SBS $j$ for uploading local model parameters ($e{_{j}^{\rm up}}$) is
\begin{equation}
	e{_{j}^{\rm up}}=p{_{j}}t{_{j}^{\rm up}}.
\end{equation}

Therefore the corresponding energy consumption $e^{\rm one}$ completing one round of model aggregation considered in this paper is given by
\begin{equation}
	e^{\rm one}=\sum_{j\in\mathcal{J}}\left({e{_{j}^{\rm cmp}}+e{_{j}^{\rm up}}}\right).
\end{equation}

\subsection{Federated Learning Model}
This section focuses on local model training at the edge server and model aggregation at the MBS.
For SBS $j$, the total number of data samples collected is:
\begin{equation}
	N_{j}=\sum_{k\in{\mathcal{S}_{j}}}N_{k}^{\rm s},
\end{equation}
Let $\mathbf{x}_{jq}$ denote an input vector of the FL algorithm in SBS $j$ and $y_{jq}$ be the corresponding output of $\textbf{x}_{jq}$, where $q\in\{1,2,\ldots,N_j^{\rm s}\}$ is the sequence number of the sample.
For each edge server $j$, the problem to be solved by local model training is $\min \limits_{\mathcal{W}_{j}}f\left(\mathcal{W}_{j},\mathbf{x}_{jq},y_{jq}\right)$.
While the training process of an FL algorithm is done in a way to solve the following optimization problem:
\begin{subequations}\label{federated_loss_minimization}
	\begin{eqnarray}
		\label{federated_loss_minimization_objective}
		&\min \limits_{ \mathcal{W}_{1},\mathcal{W}_{2},\ldots,\mathcal{W}_{J} } &   \frac{1}{\sum_{j}N_j^{\rm s}}\sum_{j=1}^J\sum_{q=1}^{N_j^{\rm s}}f\left(\mathcal{W}_{j},\mathbf{x}_{jq},y_{jq}\right)\\
		\label{federated_loss_minimization_constraint}
		&{\rm s.t.}&\mathcal{W}_{1}=\mathcal{W}_{2}=\ldots=\mathcal{W}_{J}=\mathcal{W},
	\end{eqnarray}
\end{subequations}
To solve the formulated problem (\ref{federated_loss_minimization}), for SBS $j$, it will transmit the local optimal $\mathcal{W}_{j}$ to MBS for aggregation as soon as it complete local calculations.

It is worth noting that MBS will only use those available parameters for aggregation.
The available parameters are parameters that have no errors during wireless transmission.
In fact the random channel variations will case packet error which have significant degradation effect on the global FL model accuracy.
The packet error rate during transmission over Rayleigh channel is given by \cite{xi2011general}:
\begin{equation}
	Error_{j}=\sum_{n=1}^JC_{j,n}\left(1-exp\left(\frac{-mB{N_{0}}}{p{_{j}}h_{j,n}}\right)\right),
\end{equation}
where $m$ represents the waterfall threshold.

Let $F\left(\mathcal{W} \right)$ denote the global loss function shown in formula (\ref{federated_loss_minimization_objective}).
According to \cite{chen2020joint}, we give the following remark:

\begin{remark}
	When the global loss function $F\left(\mathcal{W} \right)$ satisfies:
	\begin{itemize}
		\item [1)]
		The gradient $\nabla F\left(\mathcal{W} \right)$of  $F\left(\mathcal{W} \right)$ is uniformly Lipschitz continuous with respect to $\mathcal{W}$, that is
		\begin{equation}\label{asumption_1}
			\left\|\nabla F\left(\mathcal{W}_{t+1} \right)-\nabla F\left(\mathcal{W}_{t} \right)\right\| \leq L\left\|\mathcal{W}_{t+1}-\mathcal{W}_{t} \right\|,
		\end{equation}
		\item [2)]
		$F\left(\mathcal{W} \right)$ is strongly convex with positive parameter $\mu$:
		\begin{equation}\label{asumption_2}
			\begin{aligned}
				&F\left(\mathcal{W}_{t+1} \right) \geq F\left(\mathcal{W}_{t} \right)+\left(\mathcal{W}_{t+1}-\mathcal{W}_{t} \right)^{T}\nabla F\left(\mathcal{W}_{t} \right)\\
				&\qquad\qquad\quad  +\frac{\mu}{2}{\left\|\mathcal{W}_{t+1}-\mathcal{W}_{t}\right\|}^2,
			\end{aligned}
		\end{equation}
		\item [3)]
		$F\left(\mathcal{W} \right)$ is twice-continuously differentiable. Based on formula (\ref{asumption_1}) and (\ref{asumption_2}), the expression can be written as:
		\begin{equation}\label{asumption_3}
			\mu \mathbf{I} \preceq \nabla^{2}F\left(\mathcal{W} \right) \preceq L\mathbf{I},
		\end{equation}
		where $\mathbf{I}$ denotes the identity matrix.
		\item [4)]
		Finally, we assume that with variables $\xi_{1},\xi_{2} \geq 0$, $F\left(\mathcal{W} \right)$ satisfies the following equation.
		\begin{equation}\label{asumption_4}
			{\left\|\nabla f\left(\mathcal{W}_{t},\mathbf{x}_{jn},y_{jn}\right)\right\|}^2-\xi_{2}{\left\|\nabla F\left(\mathcal{W}_{t}\right)\right\|}^2 \leq \xi_{1},
		\end{equation}
	\end{itemize}
	Thus, the cost function $C_{\rm learn}$ that counts for the effect of packet error rate on the FL model accuracy is given by:
	\begin{equation}
		C_{\rm learn}=\sum_{j\in\mathcal{J}}D{_{j}}Error_{j},
	\end{equation}
\end{remark}

\subsection{Problem Formulation}
In the scenario considered, the latency, energy consumption and learning loss costs incurred during federated learning are all areas of focus.
To simplify the notation, we first define the system cost as
\begin{equation}
	C_{\rm system}=\alpha{t^{\rm one}}+\left(1-\alpha\right){e^{\rm one}},
\end{equation}
where $\alpha\in{[0,1]}$ is the trade-off between dealy and energy consumption, which can be considered as a system paramater.

To sum up, we further define the total cost $C_{total}$ as
\begin{equation}
	C_{\rm total}=\rho{C_{\rm system}}+\left(1-\rho\right){C_{\rm learn}},
\end{equation}
where $\rho\in{[0,1]}$ is the trade-off between system cost and learning cost, which is also a system paramater and nees to be determined jointly by all isolated organizations.

The objective of this work is to minimize the total cost $C_{\rm total}$ by optimizing the bandwidth allocation to all sensors, the CPU frequency used to train local model, the transmit power at SBSs as well as the subcarrier allocation scheme at the MBS.
As a result, the optimization problem is formulated as
\begin{subequations}\label{total_cost_minimization}
	\begin{eqnarray}
		\label{total_cost_minimization_objective}
		&\min \limits_{ \mathbf{B},\mathbf{F},\mathbf{P},\mathbf{C} } &  C_{\rm total} \\
		\label{total_cost_minimization_bandwidth1_constraint}
		&{\rm s.t.}&\sum_{k\in{\mathcal{S}_{j}}}B_{k}^{\rm s}\le B{_{j}},\ \forall{j}, \\
		\label{total_cost_minimization_bandwidth2_constraint}
		&{}&B_{k}^{\rm s}\ge 0,\ \forall{k},\\
		\label{total_cost_minimization_power_constraint}
		&{}&0 \le p_{j} \le p{_{j}^{\rm max}},\ \forall{j},\\
		\label{total_cost_minimization_subcarrier1_constraint}
		&{}&\sum_{j=1}^JC_{j,n} = 1,\ \forall{n},\\
		\label{total_cost_minimization_subcarrier2_constraint}
		&{}&\sum_{n=1}^JC_{j,n} = 1,\ \forall{j},\\
		\label{total_cost_minimization_subcarrier3_constraint}
		&{}&C_{j,n}\in{\{0,1}\},\ \forall{j},{n},\\
		\label{total_cost_minimization_frequency_constraint}
		&{}&0 \le f_{j} \le f{_{j}^{\rm max}},\ \forall{j}, 
	\end{eqnarray}
\end{subequations}
where $\mathbf{B}=\left[B_{1}^{\rm s},\ldots,B_{K}^{\rm s}\right]^T$, $\mathbf{F}=\left[f{_{1}},\ldots,f{_{J}}\right]^T$, $\mathbf{P}=\left[p{_{1}},\ldots,p{_{J}}\right]^T$,
$\mathbf{C}=\left[C_{1,1},\cdots,C_{1,n},\ldots,C_{J,J}\right]^T$. 
$B{_{j}}$, $p{_{j}^{\rm max}}$ and $f{_{j}^{\rm max}}$ are respectively the maximum bandwidth, maximum transmit power and maximum local computation capacity of SBS $j$.
(\ref{total_cost_minimization_bandwidth1_constraint}) and (\ref{total_cost_minimization_bandwidth2_constraint}) are both bandwidth constraints.
Specifically, (\ref{total_cost_minimization_bandwidth1_constraint}) means that the total bandwidth allocated by SBS $j$ to the sensors within its range does not exceed its own budget, while (\ref{total_cost_minimization_bandwidth2_constraint}) means that the bandwidth allocated to each sensor must be positive to ensure reasonable.
(\ref{total_cost_minimization_power_constraint}) and
(\ref{total_cost_minimization_frequency_constraint}) are power constraints and computing frequency constraints, respectively.
While (\ref{total_cost_minimization_subcarrier1_constraint}) to (\ref{total_cost_minimization_subcarrier3_constraint}) are the conditions to be satisfied by the feasible subcarrier allocation scheme. 

\begin{figure} [t!]
	\centering
	\includegraphics[width=2.9 in]{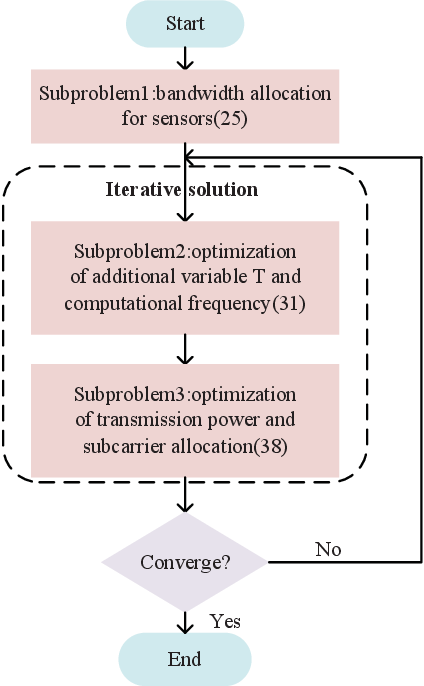}
	\caption{ Schematic diagram of problem solving process.}
	\label{problem_flow}
\end{figure}

Because of the existence of the discrete optimization variable $\mathbf{C}$ and the non-convexity of the optimization object, the problem (\ref{total_cost_minimization}) is a mixed integer nonlinear programing (MINLP) which is pretty hard to find the global optimal solution. 
Although in this problem the subcarrier selection scheme C is the only discrete variable, the exhaustive search for subcarrier allocation schemes is still infeasible.
Since the computational complexity of this method would be the factorial level.
Problem decomposing can play an important role in simplifying the problem, so we will further analyze and decompose the proposed problem (\ref{total_cost_minimization}) to find the solution method in the following.
The problem solving flow chart is shown in Figure \ref{problem_flow}.
The detailed algorithms for each subproblem will be introduced in the next section.

\section{Problem Solution}
\begin{figure} [t!]
	\centering
	\includegraphics[width=3.2 in]{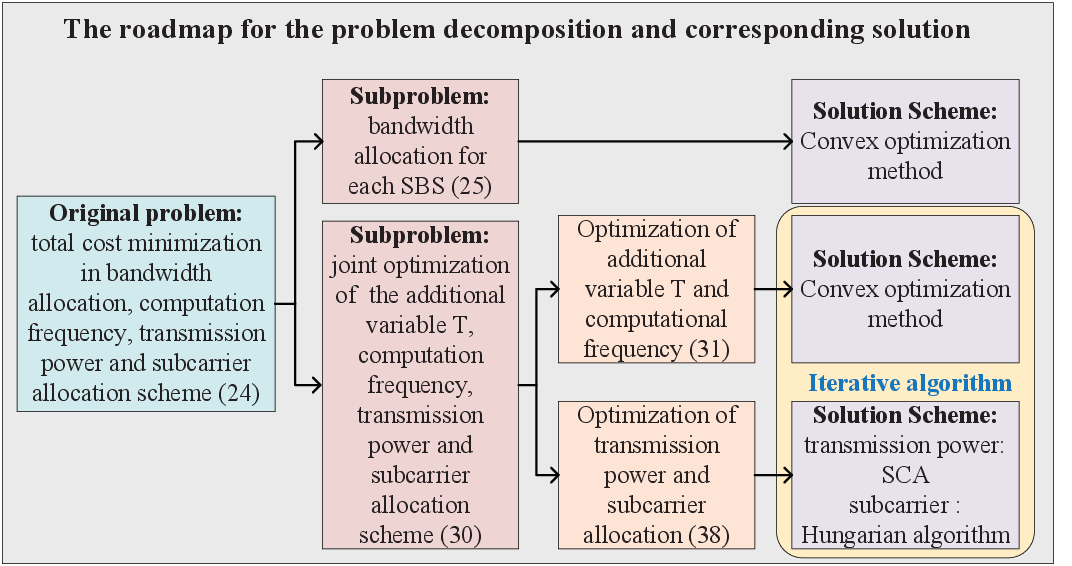}
	\caption{ The roadmap for the problem decomposition and corresponding solution.}
	\label{roadmap_problem}
\end{figure}
The overview of the roadmap for the problem decomposition and corresponding solution is shown in Figure \ref{roadmap_problem}.
It can be clearly seen from it that the original problem (\ref{total_cost_minimization}) is decomposed into several subproblems.
The closed-form solution for the bandwidth allocation subproblem can be obtained by analysis.
The remaining joint optimization problems can be further decoupled and solved by iterative optimization method.
In the follow-up part of this section, we will give a detailed description of these subproblems and the corresponding solutions designed.

\subsection{Sensor Bandwidth Optimization}
\begin{figure} [t!]
	\centering
	\includegraphics[width=3.2 in]{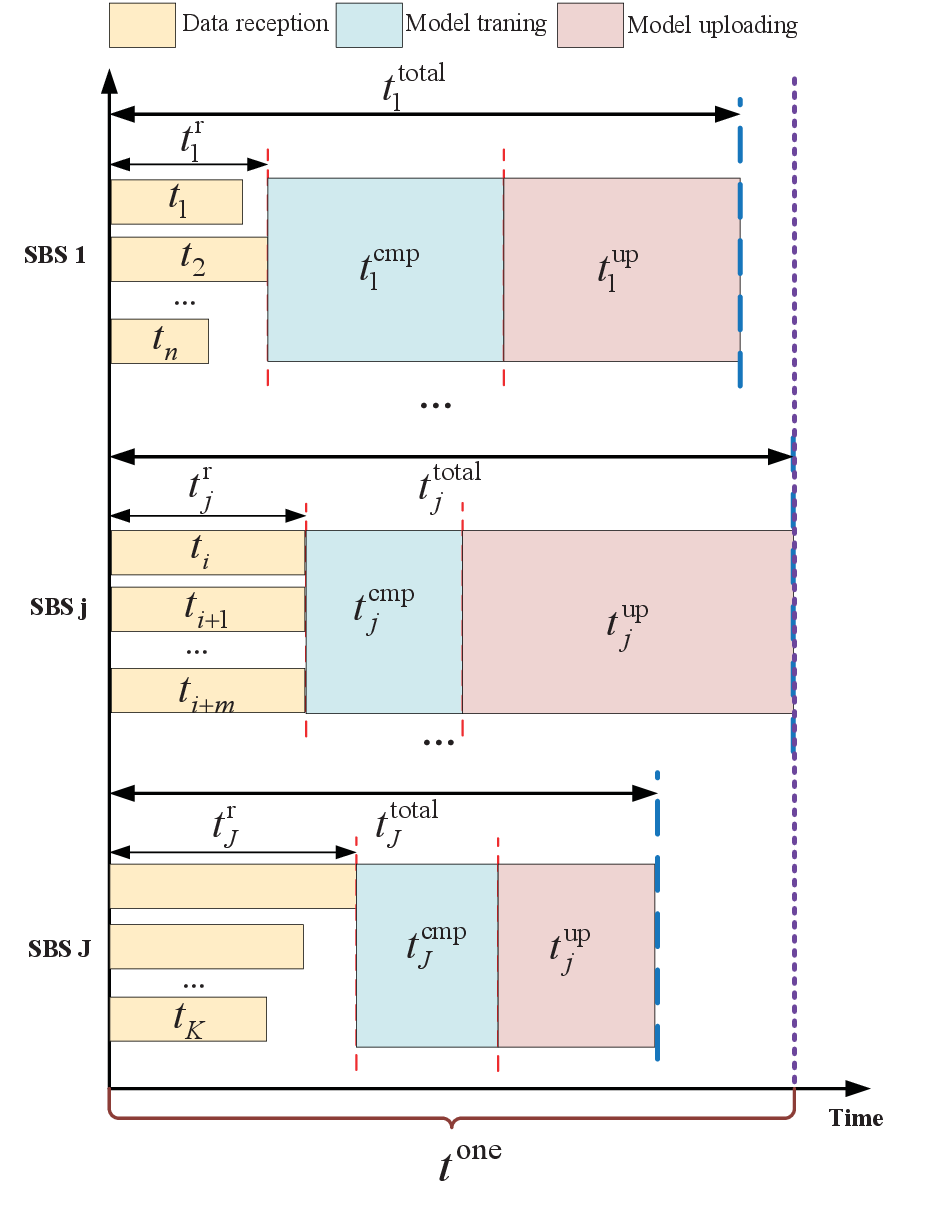}
	\caption{Diagram of federated learning flows.}
	\label{bandwidth_allocation}
\end{figure}

Fig. \ref{bandwidth_allocation} depicts a schematic composition of federated learning flows in this article.
In fact, the allocation of bandwidth to the sensor will only affect the data reception time, i.e. $t{_{j}^{\rm r}} (\forall{j})$ in the figure.
By analyzing the conditions required for $t^{\rm one}$ to reach the optimum, we can obtain the following proposition.
\begin{proposition} \label{proposition_1}
	When $t^{\rm one}$ reaches the optimal value $t{^{*}}$, for any SBS $j$, if $t{_{j}^{\rm total}}=t^*$, $t{_{j}^{\rm r}}$ minimizes and the time it takes to receive data from all sensors within its range is equal.
\end{proposition}

\begin{IEEEproof}
	Please see Appendix A.
\end{IEEEproof}

In fact, in order to shorten the time spent on federated learning, any SBS tends to spend the least amount of time receiving sensor data.
In addition, bandwidth allocation schemes of different SBSs are independent of each other, so the bandwidth optimization subproblem for SBS $j$ can be formulated as
\begin{subequations}\label{t_j_r_minimization}
	\begin{eqnarray}
		\label{t_j_r_minimization_objective}
		&\min \limits_{ \mathbf{B}_j } &  t_j^{\rm r} \\
		\label{t_j_r_constraint_1}
		&{\rm s.t.}&\sum_{k\in{\mathcal{S}_{j}}}B_{k}^{\rm s}\le B{_{j}},\\
		\label{t_j_r_constraint_2}
		&{}&B_{k}^{\rm s}\ge 0,\ \forall{k} \in{\mathcal{S}_j}, 
	\end{eqnarray}
\end{subequations}
where $\mathbf{B}_j=\{B_k^{\rm s}\},\forall{k\in{\mathcal{S}_j}}$.
Based on \textbf{Proposition \ref{proposition_1}}, we can get $t{_{j}^{r}}$ is minimized if and only if the time to receive data is equal within an organization.
As a result, the SBS can start local model training earlier to speed up the federated learning process.
Based on the above analysis, the optimal sensor bandwidth allocation for SBS $j$ should meet the following requirements:
\begin{equation} \label{equation_bandwidth}
	\left\{\begin{array}{rcl}
		\frac{D_{k}^{\rm s}}{{B_{k}^{\rm s}}^*{\rm log}_{2}\left(1+p^{\rm max}h_{k}^{\rm s}/B{_{j}}N_{0}\right)}=a{_{j}^{*}},&&{\forall{k}\in{\mathcal{S}_{j}}},\\
		\sum_{{k}\in{\mathcal{S}_{j}}}{B_{k}^{\rm s}}^*=B{_{j}},&&{}
	\end{array}\right.
\end{equation}
where $a{_{j}^{*}}$ is a positive constant representing the optimal data receiving time of SBS $j$, i.e., $t{_{j}^{\rm r}}^{*}=a{_{j}^{*}}$.
The solution of Equation (\ref{equation_bandwidth}) satisfies the following theorem
\begin{theorem}\label{theorem_1}
	The solution of the equation set (\ref{equation_bandwidth}) is
\begin{equation}\label{optimal_B}
	{B_{k}^{\rm{s}}}^*=\frac{B_jD_{k}^{\rm s}/{\rm log}_{2}\left(1+\frac{p^{\rm max}h_{k}^{\rm s}}{B{_{j}}N_{0}}\right)}{\sum_{{i}\in\mathcal{S}_{j}}D_{i}^{\rm s}/{\rm log}_{2}\left(1+\frac{p^{\rm max}h_{i}^{\rm s}}{B{_{j}}N_{0}}\right)},\ \forall{k}\in\mathcal{S}_{j},
\end{equation}
\begin{equation}\label{optimal_a}
	a{_{j}^{*}}=\frac{\sum_{{i}\in\mathcal{S}_{j}}D_{i}^{\rm s}/{\rm log}_{2}\left(1+\frac{p^{\rm max}h_{i}^{\rm s}}{B{_{j}}N_{0}}\right)}{B_j}.
\end{equation}
\end{theorem}

\begin{IEEEproof}
Please see Appendix B.
\end{IEEEproof}

%

By substituting $t{_{j}^{\rm r}}^{*}=a{_{j}^{*}}$ into problem (\ref{total_cost_minimization}), the optimization problem can be simplified as
\begin{subequations} \label{re1_total_cost_minimization}
	\begin{eqnarray}
		&\min \limits_{ \mathbf{F},\mathbf{P},\mathbf{C} } & \rho\alpha\max_{j}\left(a{_{j}^{*}}+t{_{j}^{\rm cmp}}+t{_{j}^{\rm up}}\right)+\rho\left(1-\alpha\right)e^{\rm one} \nonumber\\
		&{}&+\left(1-\rho\right)C_{\rm learn}  \\
		&{\rm s.t.}& (\ref{total_cost_minimization_power_constraint}),(\ref{total_cost_minimization_subcarrier1_constraint}),(\ref{total_cost_minimization_subcarrier2_constraint}),(\ref{total_cost_minimization_subcarrier3_constraint}),(\ref{total_cost_minimization_frequency_constraint}). 
	\end{eqnarray}
\end{subequations}
Further introducing the additional variable $T$, problem (\ref{re1_total_cost_minimization}) is equivalent to the following form:
\begin{subequations} \label{re2_total_cost_minimization}
	\begin{eqnarray}
		&\min \limits_{ \mathbf{F},\mathbf{P},\mathbf{C},\mathbf{T} } & \rho\alpha{T}+\rho\left(1-\alpha\right)e^{\rm one}+(1-\rho)C_{\rm learn}  \\
		&{\rm s.t.}& (\ref{total_cost_minimization_power_constraint}),(\ref{total_cost_minimization_subcarrier1_constraint}),(\ref{total_cost_minimization_subcarrier2_constraint}),(\ref{total_cost_minimization_subcarrier3_constraint}),(\ref{total_cost_minimization_frequency_constraint}),\\
		\label{add_variable_t}
		&{}& a{_{j}^{*}}+t{_{j}^{\rm cmp}}+t{_{j}^{\rm up}}\le{T},\ \forall{j}.
	\end{eqnarray}
\end{subequations}
In problem (\ref{re2_total_cost_minimization}), the optimization variable $\mathbf{B}$ in problem (\ref{total_cost_minimization}) is removed while the optimization variable $\mathbf{T}$ is added.
The optimization problem is still a mixed integer nonlinear programming problem.
But by analyzing the form of it, the problem can be further decomposed into two sub-problems and solved by iterative algorithm.
We will focus on this iterative algorithm in the following sections.
%

The iterative algorithm can be roughly divided into two steps in each iteration.
To optimize $(\mathbf{F},\mathbf{P},\mathbf{C},\mathbf{T})$ in problem (\ref{re2_total_cost_minimization}), we first optimize $(\mathbf{F}, \mathbf{T})$ with fixed $(\mathbf{P}, \mathbf{C})$, then $(\mathbf{P}, \mathbf{C})$ is solved based on the obtained $(\mathbf{F}, \mathbf{T})$ in the previous step.

\subsection{Latency and Computation Frequency Optimization}  
%
First, when the subcarrier allocation scheme $\mathbf{C}$ and the corresponding transmitting power $\mathbf{P}$ are given, problem (\ref{re2_total_cost_minimization} can be transformed into:
\begin{subequations} \label{T_F_minimization}
	\begin{eqnarray}
		\label{T_F_min_obj}
		&\min \limits_{ \mathbf{F},\mathbf{T} } 
		& \rho\alpha{T}+\rho\left(1-\alpha\right)\sum_{j\in\mathcal{J}}\kappa\epsilon{D{_{j}}}\left(f{_{j}}\right)^{2}\\
		\label{T_F_min_T}
		&{\rm s.t.}& T-\frac{\epsilon D{_{j}}}{f{_{j}}}\ge a{_{j}^{*}}+t{_{j}^{\rm up}},\ \forall{j}\\
		&{}& 0 \le f{_{j}} \le f{_{j}^{\rm max}},\ \forall{j}.
	\end{eqnarray}
\end{subequations}
It can be seen from (\ref{T_F_min_obj}) that it is most effective for all SBSs to use the minimal computation capacity.
Therefore, through constraint (\ref{T_F_min_T}), the expression of the minimum computation frequency of SBS $j$ can be obtained as 
\begin{equation}\label{optimal_F_temp}
	\tilde{f}_{j}^{*}=\frac{\epsilon D{_{j}}}{T-a{_{j}^{*}}-t{_{j}^{\rm up}}}.
\end{equation}
Note that $\tilde{f}_{j}^{*}$ is still a function of $T$, and we can calculate the optimal frequency as soon as we solve for $T$.
By substituting the above equation (\ref{optimal_F_temp}) into the objective function (\ref{T_F_min_obj}), the optimization problem is transformed into a univariate optimization problem with respect to $\mathbf{T}$
\begin{subequations} \label{T_minimization}
	\begin{eqnarray}
		\label{T_min_obj}
		&\min \limits_{\mathbf{T} } 
		& \rho\alpha{T}+\rho\left(1-\alpha\right)\kappa\epsilon^{3}\sum_{j\in\mathcal{J}}\frac{\left(D_{j}\right)^3}{\left(T-a_{j}^{*}-t_{j}^{\rm up}\right)^2}\\
		\label{T_min_T}
		&{\rm s.t.}& T \ge T^{\rm min}.
	\end{eqnarray}
\end{subequations}
where the definition of $T^{\rm min}$ in constraint (\ref{T_F_min_T}) is:
\begin{equation}
	T^{\rm min}\overset{\text{def}}{=}\max_{j\in\mathcal{J}}\left(a_{j}^{*}+t_{j}^{\rm up}+\frac{\epsilon D_{j}}{f_{j}^{\rm max}}\right).
\end{equation}
Through the nonnegative property of the second derivative of the objective function (\ref{T_min_obj}), the problem (\ref{T_minimization}) is a convex for $\mathbf{T}$.
By taking the first-order derivative of the target function (\ref{T_F_min_obj}), we can get the following function
\begin{equation}\label{first_order_T}
	g(T)=\rho\alpha-2\rho\left(1-\alpha\right)\kappa\epsilon^{3}\sum_{j\in\mathcal{J}}\frac{\left(D_{j}\right)^3}{\left(T-a_{j}^{*}-t_{j}^{\rm up}\right)^3},
\end{equation}
The function $g(T)$ is monotonically increasing with respect to $\mathbf{T}$.
Therefore, the optimal value of $T$ at the feasible interval $[T^{min},+\infty)$ can be given by
\begin{equation}\label{optimal_T}
	T^{*}=\left\{
	\begin{array}{rcl}
		T^{\rm min},& &{g\left(T^{\rm min}\right) \ge 0,}\\
		\hat{T},& &{\rm else,}
	\end{array}\right.
\end{equation}
where $\hat{T}$ is the solution of the equation $g(T)=0$. 
Although it is not easy to solve the equation directly, binary search method can be effectively used to find the solution.
Substituting the value of $T^{*}$ into formula (\ref{optimal_F_temp}), the optimal allocations of computing capacity for all SBSs are obtained, denoted by:
\begin{equation}\label{optimal_F}
	f_{j}^{*}=\frac{\epsilon D{_{j}}}{T^{*}-a{_{j}^{*}}-t{_{j}^{\rm up}}},\forall {j}.
\end{equation}

\subsection{Transmission Power and Subcarrier Optimization} 
%
With fixed $(\mathbf{F},\mathbf{T})$, problem (\ref{re2_total_cost_minimization}) can be simpicated to:
\begin{subequations} \label{P_C_minimization}
	\begin{eqnarray}
		\label{P_C_min_obj}
		&\min \limits_{ \mathbf{P},\mathbf{C} } 
		& \rho\left(1-\alpha\right)\sum_{j\in\mathcal{J}}\dfrac{p_{j}D}{\sum_{n}C_{j,n}B{\rm log}_{2}\left(1+p_{j}h_{j,n}/BN_{0}\right)} \nonumber\\
		& 
		&+\left(1-\rho\right)\sum_{j\in\mathcal{J}}\left(D_{j}Error_{j}\right)\\
		\label{P_C_min_constraint}
		&{\rm s.t.}& (\ref{total_cost_minimization_power_constraint}),(\ref{total_cost_minimization_subcarrier1_constraint}),(\ref{total_cost_minimization_subcarrier2_constraint}),(\ref{total_cost_minimization_subcarrier3_constraint}),(\ref{add_variable_t}).
	\end{eqnarray}
\end{subequations}
In this problem (\ref{P_C_min_constraint}), the variable $\mathbf{C}$ is a discrete variable, which is more difficult to deal with than the continuous variable $\mathbf{P}$.
Therefore, we assume that the subcarrier allocation scheme has been given, and first consider the power optimization problem in this case.
Let the function $m(j)$ denotes the mapping function from SBS $j$ to the subcarrier assigned to it.
Therefore, the expression of power optimization problem is 
\begin{subequations} \label{P_minimization}
	\begin{eqnarray}
		\label{P_min_obj}
		&\min \limits_{ \mathbf{P} } 
		& \rho\left(1-\alpha\right)\sum_{j\in\mathcal{J}}\frac{p_{j}D}{B{\rm log}_{2}\left(1+p_{j}h_{j,m\left(j\right)}/BN_{0}\right)}+ \nonumber\\
		& &\left(1-\rho\right)\sum_{j\in\mathcal{J}}\left(D_{j}\left(1-exp\left(\frac{-mBN_{0}}{p_{j}h_{j,m(j)}}\right)\right)\right) \nonumber \\
		&{}&{}\\
		\label{P_min_constraint1}
		&{\rm s.t.}&\frac{D}{B{\rm log}_{2}\left(1+\frac{p_{j}h_{j,m\left(j\right)}}{BN_{0}}\right)} \le T-a_{j}^{*}-t_{j}^{\rm cmp},\ \forall{j} \nonumber\\
		&{}&{}\\
		\label{P_min_constraint2}
		&{}&0 \le p_{j} \le p_{j}^{\rm max},\ \forall{j}.
	\end{eqnarray}
\end{subequations}
The expression of formula (\ref{P_min_obj}) is the sum of the transmission energy consumption and learning cost of all organizations. 
In fact, the optimal transmission power of all SBSs is not coupled in problem (\ref{P_minimization}).
Therefore, the optimization problem can be decomposed into $J$ unrelated optimization problems, that is, each problem is an optimization problem for the transmission power of one SBS.
Take SBS $j$ as an example, its optimization problem is shown as
\begin{subequations} \label{isolated_P_minimization}
	\begin{eqnarray}
		\label{isolated_P_min_obj}
		&\min \limits_{ p_{j} } 
		& \rho\left(1-\alpha\right)\frac{p_{j}D}{B{\rm log}_{2}\left(1+p_{j}h_{j,m\left(j\right)}/BN_{0}\right)} \nonumber\\
		& &+\left(1-\rho\right)D_{j}\left(1-exp\left(\frac{-mBN_{0}}{p_{j}h_{j,m(j)}}\right)\right)  \nonumber\\
		&{}&{}\\
		\label{isolated_P_min_constraint1}
		&{\rm s.t.}&p_{j}^{\rm min} \le p_{j} \le p_{j}^{\rm max}.
	\end{eqnarray}
\end{subequations}
where $p_{j}^{\rm min}$ in (\ref{isolated_P_min_constraint1}) is transformed from constraint (\ref{P_min_constraint1}) in problem (\ref{P_minimization}) and is given by
\begin{equation}\label{p_j_min}
	p_{j}^{\rm min}=\frac{BN_{0}}{h_{j,m\left(j\right)}}\left(2^{\frac{D}{B\left(T-a_{j}^{*}-\epsilon D_{j}/f_{j}\right)}}-1\right).
\end{equation}
It should be noted that when $p_{j}^{\rm min}$ is calculated to be greater than $p_{j}^{\rm max}$ on a certain subcarrier, it indicates that the selection of this subcarrier cannot achieve the expected latency.
Therefore, the cost of selecting this subcarrier needs to be assigned a large positive number to avoid selecting it.
The problem (\ref{isolated_P_minimization}) is an univariate nonconvex optimization problem.
The successive convex approximation (SCA) algorithm can be used to solve this problem.
To simplify notation, denote formula (\ref{isolated_P_min_obj}) as $h(p_{j})$.
In the $n$-th iteration of the successive convex approximation algorithm, we take the function $g(p_{j},p_{j}^{n-1})$ as the approximation of the function $h(p_{j})$.
The function $g(p_{j},p_{j}^{n-1})$ is defined as
\begin{equation}\label{SCA_approach_function}
	\begin{aligned}
		&g\left(p_{j},p_{j}^{n-1}\right)=h\left(p_{j}^{n-1}\right)+h^{'}\left(p_{j}^{n-1}\right)\left(p_{j}-p_{j}^{n-1}\right)\\
		&\qquad\qquad\qquad+\frac{\tau{_{j}}}{2}\left(p_{j}-p_{j}^{n-1}\right)^{2},
	\end{aligned}
\end{equation}
where $h^{'}(p_{j}^{n-1})$ is the function value of the first-order derivative of $h(p_{j})$ at the point $p_{j}^{n-1}$, and its expression is
\begin{equation}
	\begin{aligned}
		&h^{'}\left(p_j^{n-1}\right)=\frac{alog_{2}\left(1+bp_{j}^{n-1}\right)-abp_{j}^{n-1}/\left(\left(1+bp_{j}^{n-1}\right){\rm ln}2\right)}{{\rm log}_{2}^{2}\left(1+bp_{j}^{n-1}\right)}\\
		&\qquad\qquad\qquad-\frac{cd}{(p_j^{n-1})^2}exp\left(\frac{-d}{p_j^{n-1}}\right),
	\end{aligned}
\end{equation}
where $a=\frac{\rho\left(1-\alpha\right)D}{B}$, $b=\frac{h_{j,m\left(j\right)}}{BN_{0}}$, $c=\left(1-\rho \right)D_j$, and
$d=\frac{mBN_{0}}{h_{j,m \left(j \right)}}$.
The first two terms in formula (\ref{SCA_approach_function}) are the first-order Taylor expansion of $h\left(p_j\right)$ at $p_j^{n-1}$.
While the third term is used to ensure that $g\left(p_{j},p_{j}^{n-1}\right)$ is strongly convex, where $\tau_{j}$ is any positive constant.
Therefore the SCA-based power optimization algorithm is summarized in \textbf{Algorithm \ref{SCA_power_algorithm}}.

\begin{algorithm}[t!]
	\caption{SCA-Based Power Optimization Algorithm}
	\label{SCA_power_algorithm}
	\begin{algorithmic}[1]
		\renewcommand{\algorithmicrequire}{\textbf{Initialize}}
		\renewcommand{\algorithmicensure}{\textbf{Output}}
		\STATE \textbf{Initialize} the tolerance $\delta$, maximum iteration number $N_{1}$, and set the current iteration number as $n_1=0$.
		\STATE Find a feasible point $p_{i}^{0}\in \mathcal{P}$;
		\REPEAT
		\STATE Set $n_{1}\leftarrow n_{1}+1$;
		\STATE Let $p_j^{n_1}=\mathop{\arg\min}\limits_{p_{j} \in \mathcal{P}}g     \left(p_j,p_j^{n_1-1} \right)$;
		\STATE Set $d_{n_1}=p_j^{n_1}-p_j^{n_1-1}$;
		\STATE Armijo step-size rule: Choose $\beta\in \left(0,1 \right)$, $\sigma \in \left(0,0.5 \right)$. Let $\gamma_k=\beta^{m_k}$, where $m_k$ is the smallest non-negative integer satisfying: 
		$h\left(p_j^{n_1-1}+\beta^{m}d_{n_1} \right) \le h\left(p_j^{n_1-1}\right)+\sigma\beta^{m}h^{'}\left(p_j^{n_1-1}\right)d_{n_1}$;
		\STATE Update $p_j^{n_1}=p_j^{n_1-1}+\gamma_k{d_{n_1}}$;
		\UNTIL $|p_j^{n_1}-p_j^{n_1 - 1}| < \delta$ or $n_{1} > N_{1}$;
		\STATE \textbf{Output} the optimal $p_j^{*}$ or the converged solution $p_j^{n_1}$. 
	\end{algorithmic}
\end{algorithm}

Next we focus on the subcarrier allocation subproblem.
Regardless of the constraints (\ref{total_cost_minimization_subcarrier2_constraint}) on subcarrier allocation, each SBS can choose one of all subcarriers.
Therefore, all organizations can solve the optimal power under each subcarrier one by one through \textbf{Algorithm \ref{SCA_power_algorithm}}.
The optimal power of organization $j$ on the $n$-th subcarrier can be denoted as $p_{j,n}^{*}$.
By substituting ${p_{j,n}^{*}}$ into function $h\left(p_j\right)$, we can obtain the minimum cost value $h\left(p_{j,n}^{*}\right)$ that can be achieved for organization $j$ selecting this subcarrier.
So the original problem (\ref{P_C_minimization}) can now be converted to
\begin{subequations} \label{isolated_C_minimization}
	\begin{eqnarray}
		\label{isolated_C_min_obj}
		&\min \limits_{ \mathbf{C} } 
		& \sum_{j=1}^J\sum_{n=1}^JC_{j,n}h\left(p_{j,n}^{*}\right)\\
		\label{isolated_C_min_constraint}
		&{\rm s.t.}&(\ref{total_cost_minimization_subcarrier1_constraint}),(\ref{total_cost_minimization_subcarrier2_constraint}),(\ref{total_cost_minimization_subcarrier3_constraint}).
	\end{eqnarray}
\end{subequations}
Note that $h\left(p_{j,n}^{*}\right)(\forall{j,n})$ are always greater than 0.
Therefore, the above optimization problem (\ref{isolated_C_minimization}) can be regarded as an assignment problem and solved by the Hungarian algorithm.

So far, all the subproblems have been solved effectively.
Finally, the complete algorithm flow for problem (\ref{total_cost_minimization}) is summarized in \textbf{Algorithm \ref{complete_algorithm}}.
Next, we will further study the properties of the proposed algorithm.

\subsection{Convergence and Complexity Analysis}
\begin{proposition}\label{Convergence Analysis}
	(Convergence): The algorithm shown in \textbf{Algorithm \ref{complete_algorithm}} can always converge to a stable solution.	
\end{proposition}

\begin{IEEEproof}
	For the sensor bandwidth allocation subproblem, we can easily obtain a closed solution.
	Therefore, the convergence of our algorithm mainly depends on the properties of the iterative algorithm for solving problem (\ref{re2_total_cost_minimization}).
	In one iteration, subproblems (\ref{T_F_minimization}) and (\ref{P_C_minimization}) are solved successively.
	Given the subcarrier allocation scheme $\mathbf{C}$ and corresponding power distribution scheme $\mathbf{P}$, subproblem (\ref{T_F_minimization}) can be reformulate to a convex problem (\ref{T_minimization}).
	Thus, the optimal solution $\mathbf{T}^*$ and $\mathbf{F}^*$ are obtained in this step.

	Then with computed $\mathbf{T}^*$ and $\mathbf{F}^*$, we start to solve the other subproblem (\ref{P_C_minimization}).
	First of all, for the convenience of interpretation, we assume that the subcarrier allocation scheme and the corresponding transmission power solved in the last iteration are denoted as $c_j^{\rm last}$  and $p_j^{\rm last}$ for SBS $j$ respectively.
	Note that $c_j^{\rm last}$  and $p_j^{\rm last}$ are still included in the current feasible set.
	When using SCA method to find the optimal transmission power under each subcarrier for each SBS, $p_j^{\rm last}$ is used as the initial search point when searching in subcarrier $c_j^{\rm last}$.
	While for other subcarriers except $c_j^{\rm last}$, we can search from any feasible value.
	By doing so, we can see that even though we did not adjust the subcarrier allocation mechanism in this iteration, that is $c_j^{\rm new}=c_j^{\rm last}$, the cost value obtained in this iteration is still no higher than that in the last iteration.
	Then, the Hungarian algorithm which can give the optimal solution is applied to solve the subcarrier allocation problem (\ref{isolated_C_minimization}).
	Therefore the cost we get is the lowest, and obviously no larger than the cost under the subcarrier allocation scheme $c_j^{\rm last}$ calculated in the last iteration. 

	Based on the above analysis, we can find that the objective value of problem (\ref{re2_total_cost_minimization}) obtained by applying the proposed algorithm shown in \textbf{Algorithm \ref{complete_algorithm}} is non-increasing in each step.
	Moreover, the objective value of problem (\ref{re2_total_cost_minimization}) is lower bounded by zero. 
	Thus, our proposed algorithm  always converges to a convergent solution, which completes the proof.  
	
\end{IEEEproof}

\begin{proposition}\label{Complexity Analysis}
	(Complexity): The algorithm shown in \textbf{Algorithm \ref{complete_algorithm}} can can be thought of as consisting of the bandwidth allocation algorithm in the second line and an iterative algorithm that runs repeatedly between the third and ninth line.	
	The time complexity of the bandwidth allocation algorithm is $\mathcal{O}\left(K\right)$, while the complexity of iteration algorithm for one iteration is $\mathcal{O}\left(J^3+J^2\left(1+L_{SCA}\right)+J\left(1+{\rm log}_{2}\left(1/\delta \right)\right)\right)$.
\end{proposition}
\begin{IEEEproof}
	The second line of \textbf{Algorithm \ref{complete_algorithm}} represents the bandwidth allocation algorithm.
	The optimal bandwidth allocation of each sensor can be directly calculated through the formula (\ref{optimal_B}), so the time complexity of finding the optimal bandwidth allocation scheme for all sensors is $\mathcal{O}\left(K\right)$.

	For the iterative algorithm part shown from the third to the ninth lines, the time complexity of solving $\mathbf{T}$ by applying binary search algorithm in the fourth line is $\mathcal{O}\left(J{\rm log}_{2}\left(1/\delta \right)\right)$, where $\delta$ denotes the solution accuracy.
	Then $\mathcal{F}$ can be calculated directly using formula (\ref{optimal_F}) with the complexity of $\mathcal{O}\left(J\right)$.

	The complexity of \textbf{Algorithm \ref{SCA_power_algorithm}} applied in line 5 is $\mathcal{O}\left(L_{SCA}\right)$ \cite{boyd2004convex}, where $L_{SCA}$ is the total number of iterations within the SCA algorithm.
	Since each organization needs to apply \textbf{Algorithm \ref{SCA_power_algorithm}} on each subcarrier, the total time complexity is $\mathcal{O}\left(J^2L_{SCA}\right)$.
	Then the $h\left(p_{j,n}^*\right)$ value of each organization on each subcarrier needs to be calculated in the sixth line, so a total of $J^2$ values need to be calculated, that is, the complexity is $\mathcal{O}\left(J^2\right)$.
	Finally, the Hungarian algorithm applied in the seventh line has a time complexity of $O(J^3)$ in the worst case.
	Thus, the complexity of iteration algorithm for one iteration can be calculated as $\mathcal{O}\left(J^3+J^2\left(1+L_{SCA}\right)+J\left(1+{\rm log}_{2}\left(1/\delta \right)\right)\right)$, which completes the proof.
	
\end{IEEEproof}

\begin{algorithm}[t!]
	\caption{Joint Resource Optimization Algorithm}
	\label{complete_algorithm}
	\begin{algorithmic}[1]
		\renewcommand{\algorithmicrequire}{\textbf{Initialize}}
		\renewcommand{\algorithmicensure}{\textbf{Output}}
		\STATE \textbf{Initialize} the carrier allocation scheme $C_{0}$ and the corresponding transmitted power $P_0$.
		\STATE Calculate $B_k^{\rm s*}$ and $a_j^*$ according to equations (\ref{optimal_B}) and (\ref{optimal_a});
		\REPEAT
		\STATE Calculate $T$ and $F$ according to equations (\ref{optimal_T}) and (\ref{optimal_F}) with fixed $C$ and $P$;
		\STATE Each organization applies Algorithm \ref{SCA_power_algorithm} to compute $p_{j,n}^*$ under each subcarrier condition;
		\STATE Calculate $h\left(p_{j,n}^*\right),\forall{j,n}$;
		\STATE Apply the Hungarian algorithm to solve the current optimal subcarrier allocation scheme, then update $C$;
		\STATE Update the corresponding optimal power $P$ according to $C$;
		\UNTIL the convergence condition is satisfied;
		\STATE \textbf{Output} $\mathbf{B}^*$,$\mathbf{C}^*$,$\mathbf{P}^*$,$\mathbf{T}^*$,$\mathbf{F}^*$. 
	\end{algorithmic}
\end{algorithm}

\section{Numerical Results}
\begin{figure} [t!]
	\centering
	\includegraphics[width=2.5 in]{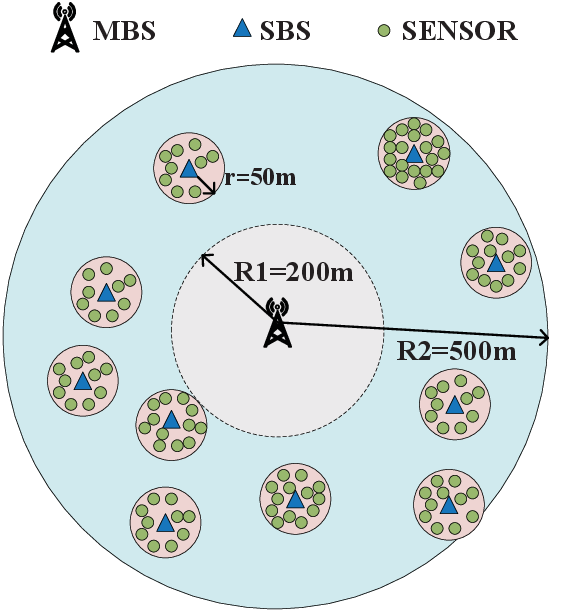}
	\caption{Sample simulation scenario.}
	\label{simulation_scenario}
\end{figure}
The sample simulation is shown in Fig.\ref{simulation_scenario}.
For our simulations, we considered a circlar area with a radius of 500m. 
One macro base station is located in the center of the area and 10 isolated organizations are uniformly distributed in the range of 200 to 500 meters from the MBS.
Each SBS occupies a circular area with a radius of 50 meters and uses one subcarrier to access to MBS. 
Each subcarrier has the same bandwidth and a total bandwidth of 3.125MHz.
In each organization, there are 10 to 20 sensors uniformly distributed within a range of 5 to 50 meters from the SBS. 
In this simulation, the positions of all SBSs and sensors are fixed.
Each time the sensor transmits 3Mbit data to the organization's SBS deployed with MEC server, and after training, the learning model with 100Kbit data size is uploaded to the MBS.
The maximum transmission power of the sensor and SBSs is 23dBm and 37dBm respectively, and the maximum computing capacity of MEC server is set 5GHz. 
For other simulation parameters, the noise power spectrum density, effective switched capacitance and processing density of learning task are set -174dBm/Hz, $2\times{10^{-29}}$ and 30 cycles/bit, respectively.

In order to validate the effectiveness of our proposed algorithms, the following five schemes are considered as benchmarks:
1) Equal bandwidth allocation algorithm: All sensors share the bandwidth resources of their SBS equally;
2) Learning cost guaranteed algorithm: The optimization objective of this algorithm only considers the learning cost (that is, $\rho$ is set to 0). 
It is worth mentioning that the optimal solution of the algorithm is always satisfied, which indicates that at least one of all organizations will calculate at the maximum calculation frequency.
After the completion of optimization, the obtained solution is used to calculate the weighted sum of system cost and learning cost; 
3) Greedy subcarrier allocation algorithm: This algorithm directly uses the Hungarian algorithm to select the channel that can maximize the sum of channel gains on the channel response matrix when selecting subcarriers; 
4) System cost guaranteed algorithm: It is similar to the second comparison algorithm, except that $\rho$ is set close to 1 when doing optimization; 
5)Time-biased algorithm: The algorithm always calculates at the maximum frequency and uses the maximum transmit power to transmit parameters, so the algorithm only optimizes the subcarrier selection. 
Next, we study the effects of various parameters such as SBS bandwidth and SBS number, on the performance of the algorithm.
During the study of the effect of each parameter (SBS bandwidth, SBS number, etc.), the number of sensors in each SBS is randomly generated from 10 to 20 and keep constant during the simulation of this parameter.
We simulate 1500 trials, and all results are averaged over independent channel realization.

\subsection{Convergence}
In this part, we simulate the convergence performance of the proposed algorithm to observe the convergence performance more intuitively and further verify \textbf{Proposition \ref{Convergence Analysis}}.
\begin{figure} [t!]
\centering
\includegraphics[width=3.2 in]{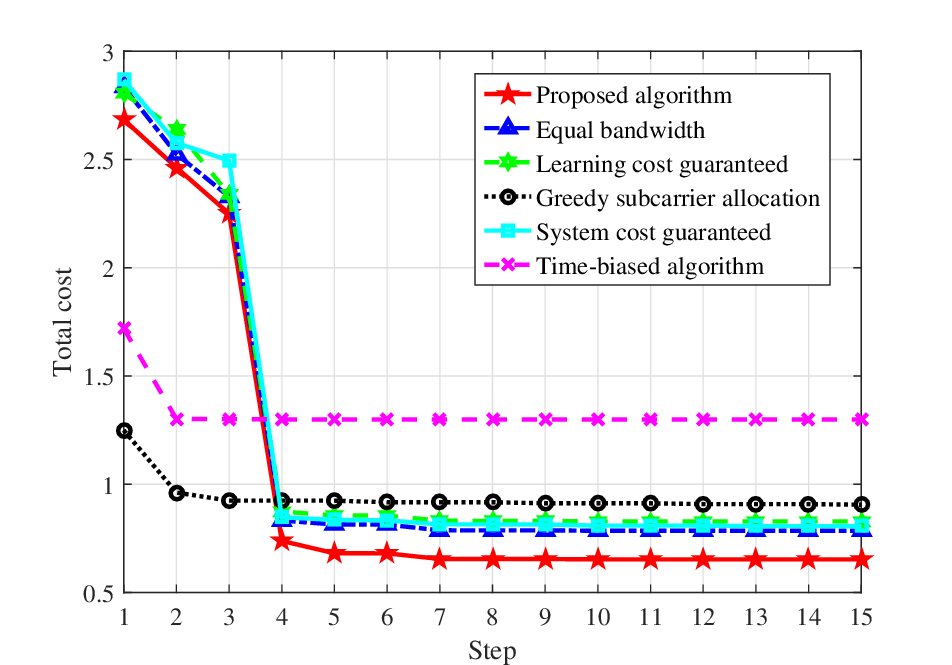}
\caption{Total cost w.r.t. step number.}
\label{convergence}
\end{figure}
Fig.\ref{convergence} shows the convergence performance of the proposed algorithm and each comparison algorithm.
It can be seen from the simulation results that all six other algorithms can converge to a stable value within 10 steps.
Since the Greedy subcarrier allocation algorithm uses a determined subcarrier allocation scheme, the Time-biased algorithm always calculates at a fixed frequency and transmits at a fixed power, so the convergence rate of the two algorithms is faster.
In addition, it is also clear from the figure that the proposed in this paper can converge quickly to the lowest total cost.

\subsection{Impact of system parameters}
\begin{figure}[t!]
\centering
	\centering
	\includegraphics[width=3.2 in]{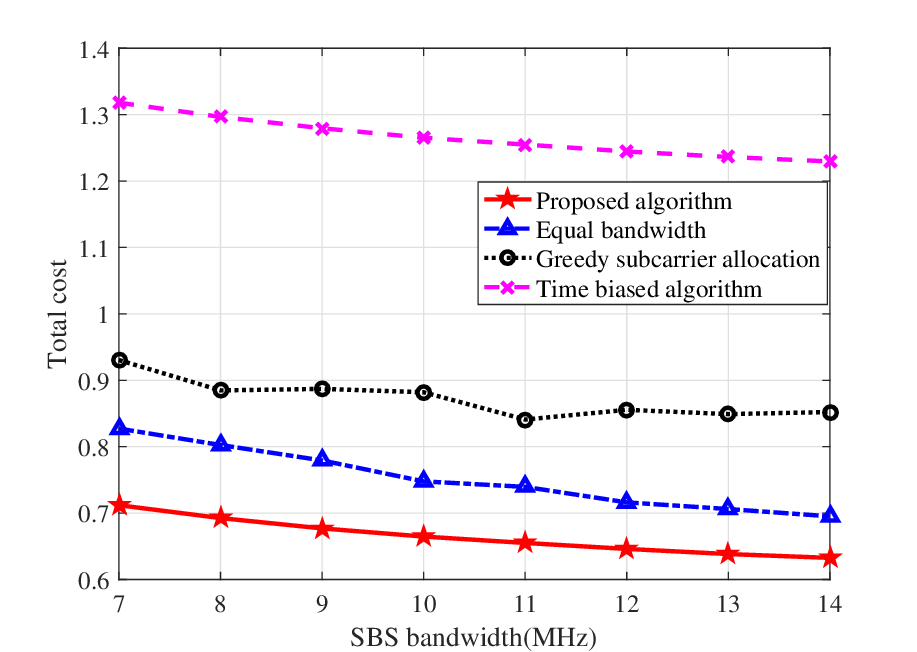}
	\caption{Total cost w.r.t. SBS bandwidth.}
	\label{SBS_bandwidth}
\end{figure}
\begin{figure}[t!]
	\centering
	\includegraphics[width=3.2 in]{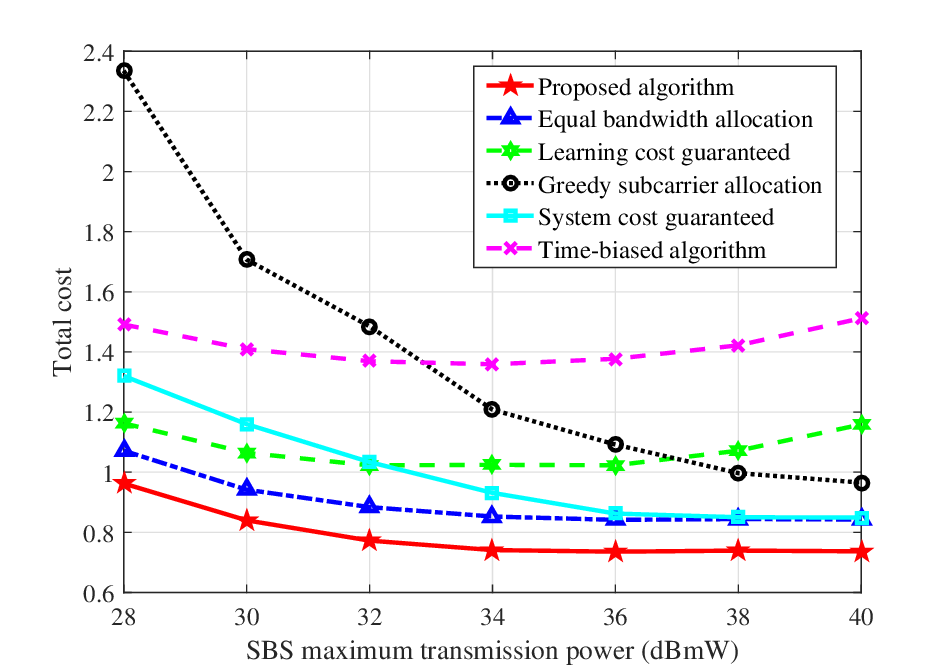}
	\caption{Total cost w.r.t. SBS maximum transmission power.}
	\label{SBS_power}
\end{figure}

\begin{figure}[t!]
	\centering
	\centering
	\includegraphics[width=3.2 in]{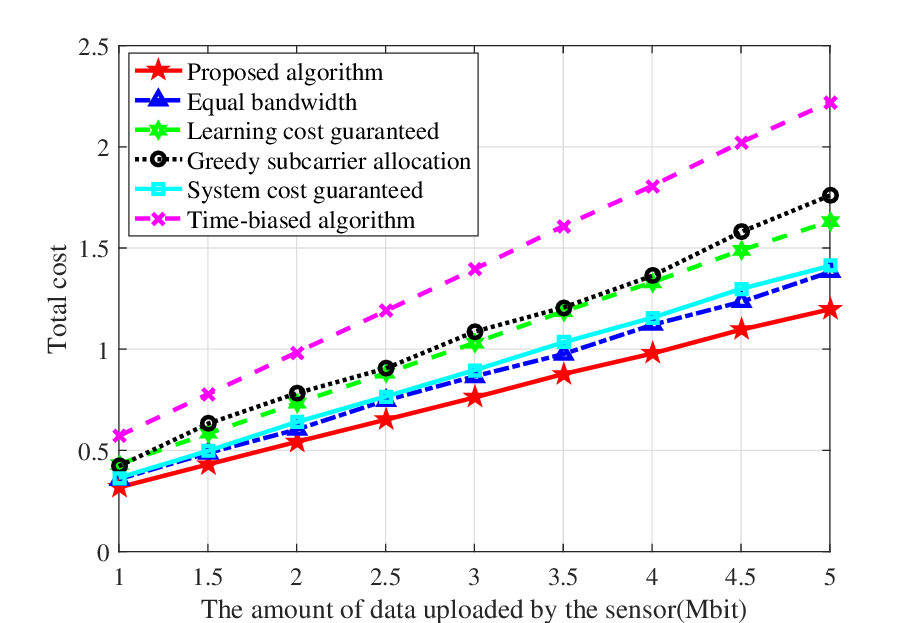}
	\caption{Total cost w.r.t. The amount of data uploaded by the sensor.}
	\label{sensor_data}
\end{figure}
\begin{figure}[t!]
	\centering
	\includegraphics[width=3.2 in]{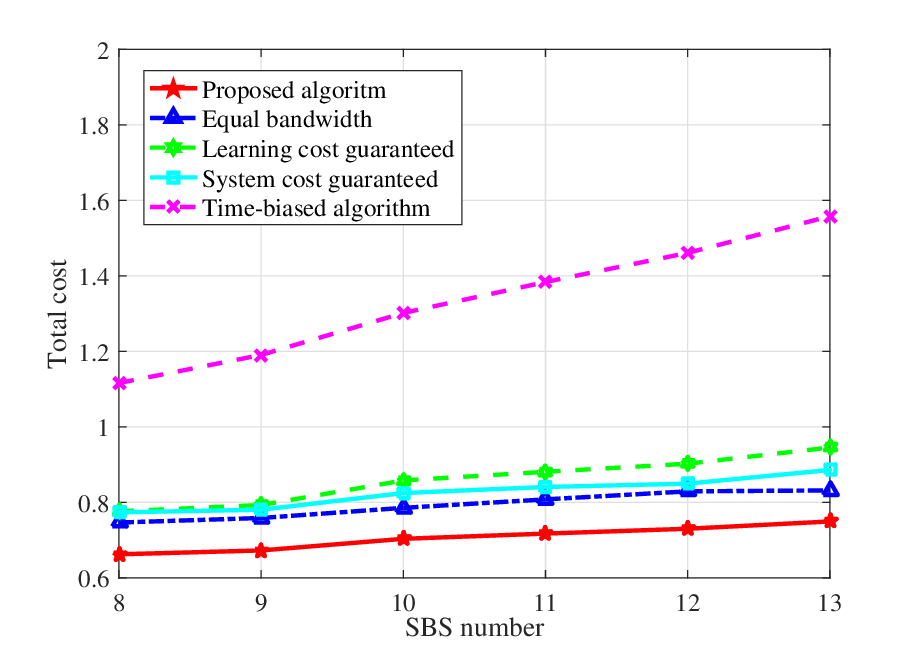}
	\caption{Total cost w.r.t. SBS number.}
	\label{organization_number}
\end{figure}
\begin{figure}[t!]
\centering
	\centering
	\includegraphics[width=3.2 in]{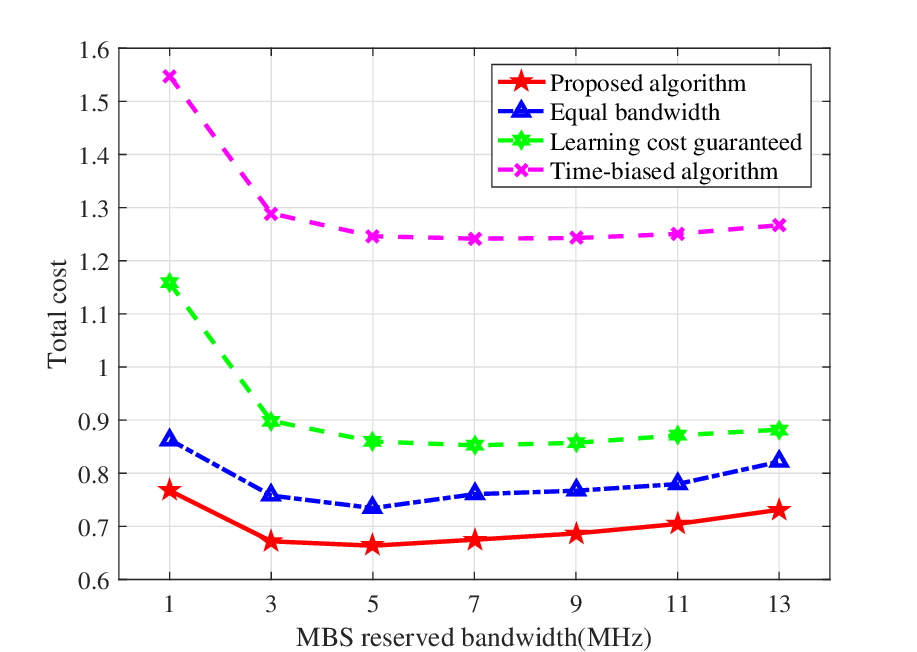}
	\caption{Total cost w.r.t. MBS reserved bandwidth.}
	\label{MBS_bandwidth}
\end{figure}
\begin{figure}[t!]
	\centering
	\includegraphics[width=3.2 in]{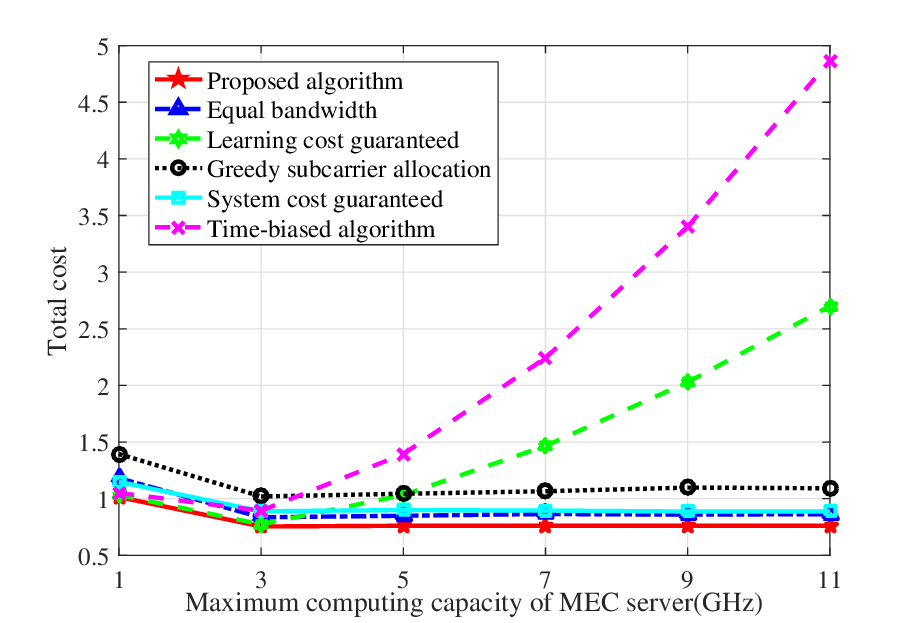}
	\caption{Total cost w.r.t. Maximum computing capacity of MEC server.}
	\label{SBS_computation}
\end{figure}

In this part, we study the performance of the proposed algorithm under different parameters.
Fig.\ref{SBS_bandwidth} shows the simulation results of each algorithm under different SBS bandwidth.
It can be seen that with the increase of SBS bandwidth, the latency term in the system cost is improved, so the total cost obtained by each algorithm shows a downward trend.
Further comparison between the proposed algorithm and the equivalent bandwidth allocation algorithm shows that: the smaller the SBS bandwidth is, the more obvious the algorithm performance gain can be brought by the SBS bandwidth optimization.
The reason for the above situation is that the sensor data upload time increases rapidly due to the reduced SBS bandwidth, so the impact of delay performance on the total cost becomes significant.
The proposed algorithm can effectively reduce the delay of data uploading through bandwidth optimization, so it can bring more obvious performance gain.

Fig.\ref{SBS_power} depicts the influence of the SBS maximum transmission power on the performance of each algorithm.
Specifically, with the increase of the maximum transmission power of SBS, the total cost achieved by the proposed algorithm, the equal bandwidth allocation algorithm and the system cost guaranteed algorithm decreases gradually at first and then stays almost constant.
The reason for this phenomenon is that when the maximum transmission power of SBS is low, the optimal power with the lowest total cost is obtained at the maximum power.
However, as the constraint of the maximum power is relaxed, the searched power value is  almost unaffected by it, so the corresponding total cost stays approximately the same.
For learning costs guaranteed algorithm and time-biased algorithm, they always take the maximum power as their optimal power choice. 
Accordingly with the increase of SBS maximum power, latency and learning cost are improved but at the same time the cost of energy consumption is worsened.
Therefore these two algorithms presented the first down and then gradually rising trend.
Finally for the greedy subcarrier allocation algorithm, due to its carrier chooses not to consider any cost performance, the latency and learning performance may be very poor.
So the algorithm has a great probability in the need to maximum power emission in some carrier.
The increase of the maximum transmitting power of SBS can more effectively make up for the loss of time delay and learning performance caused by greedy subcarrier selection, so the performance of the algorithm is improved.

Fig.\ref{sensor_data} shows the total cost as a function of the amount of data uploaded by the sensor.
It can be clearly seen from the simulation results that with the increase of the amount of data uploaded by the sensor, the total cost value obtained by each algorithm approximately presents a linear upward trend.
Because with the increase of the amount of data uploaded by the sensor, the delay, energy consumption and learning cost of the total cost will increase, so the total cost will gradually increase.
It is worth noting that the proposed algorithm always has the lowest cost value and is less affected by the amount of data uploaded by the sensor than those comparison schemes due to its joint optimization of bandwidth, computation frequency, transmission power, and subcarrier selection.
Fig.\ref{organization_number} depicts the effect of the change in the number of organizations (SBSs) on the algorithm performance.
It should be noted that in order to focus on the effect of the number of tissues, the total bandwidth of the simulated MBS is also increasing so that the bandwidth of the single carrier remains constant.
Obviously, with the increase of the number of organizations, the new organizations will bring additional energy consumption, learning costs and the possibility of further slowing down the communication time, so the total cost presents a trend of gradual increase.
In addition, by comparing the simulation results of learning cost guaranteed algorithm and system cost guaranteed algorithm, it can be found that the importance of system cost in total cost gradually increases with the increase of the number of organizations.
Considering both system cost and learning cost, the proposed algorithm always has the best performance in the case of different number of organizations.

Fig.\ref{MBS_bandwidth} compares the total cost performance of the four algorithms under different MBS reserved bandwidth.
The total cost of all algorithms generally decreases first and then increases with the increase of MBS reserved bandwidth.
When the reserved bandwidth of SBS is very small, the parameter data sent from SBS needs to experience a long latency, so the delay in the system cost becomes the main factor restricting the total cost.
With the increase of the reserved bandwidth of MBS, the delay performance is improved and the total cost decreases.
However, with the further increase of bandwidth, excessive bandwidth will cause the increase of noise power, which will deteriorate the SNR performance.
The decline of signal-to-noise ratio leads to the increase of learning cost, so the total cost starts to rise.

The simulation results shown in Fig.\ref{SBS_computation} are the impact of the maximum computing capacity of the MEC server on the performance of each algorithm.
Except for the learning cost guaranteed algorithm and the time-biased algorithm, the total cost of the other algorithms decreases gradually and then stays the same with the increase of the maximum computing capacity of MEC server.
Due to the high computing delay caused by the lack of MEC computing capacity, the delay has become a key factor limiting the total cost.
As the maximum computing capacity of MEC exceeds a certain threshold, the optimal computing frequency is independent of the maximum computing frequency, so the total cost achieved by these algorithms stay the same.
For the other two algorithms, when the maximum computing capacity of MEC server is low, the algorithm performance of the two algorithms is close to that of the proposed algorithm. 
This further confirm the analysis that the delay becomes the dominant factor in the total cost when the computing power is insufficient.
However, since the choice of computing capacity of these two algorithms is always related to the maximum computing power, when the maximum computing power exceeds a certain threshold, the performance of the algorithms deteriorates rapidly because of the rapid increase of energy consumption.

\subsection{Impact of weight coefficient}
\begin{figure} [t!]
\centering
\includegraphics[width=3.2 in]{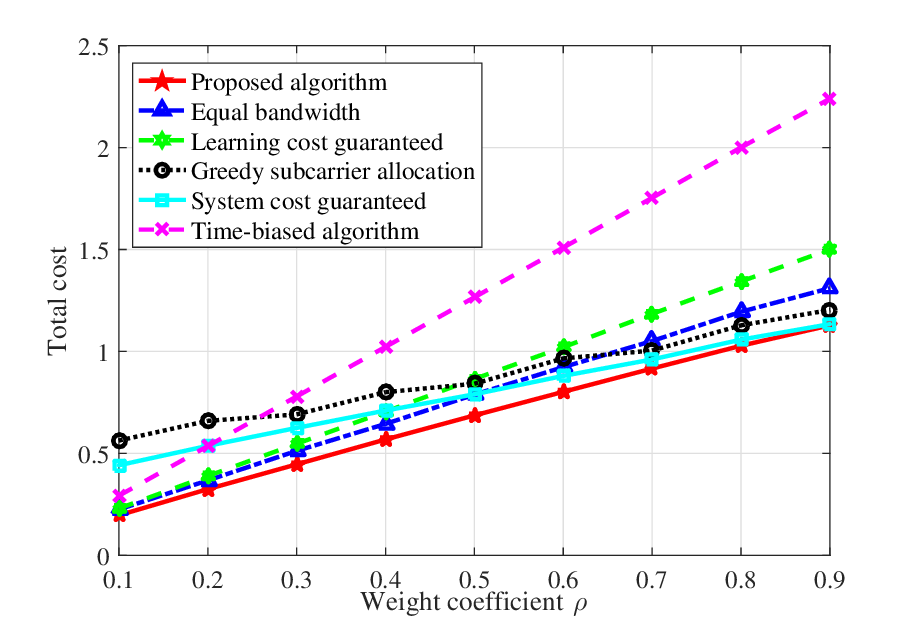}
\caption{Total cost w.r.t. Weight coefficient $\rho$.}
\label{weight}
\end{figure}
Fig.\ref{weight} shows the simulation results of algorithm performance under different weight coefficients.
It can be seen from the simulation results that when $\rho$ is low (more attention is paid to the learning cost), the performance of the proposed algorithm is very close to that of the learning cost guaranteed algorithm.
With the gradual increase of $\rho$, more attention is paid to the system cost in the total cost, so the performance of the proposed algorithm gradually approaches that of the system cost guaranteed algorithm.
In conclusion, among all schemes, the proposed algorithm can always achieve best performance through reasonable resource optimization under each weight value.

In this section, through the simulation results of the convergence of the proposed algorithm, we can see that the proposed algorithm can quickly converge to a stable value, which is consistent with the theoretical analysis in \textbf{Proposition \ref{Convergence Analysis}}. 
In addition, combining the complexity analysis in the previous section, we can clearly see the superior performance of the proposed algorithm in complexity.
As can be seen from the simulation results of the algorithms under different parameter settings in Fig. \ref{SBS_bandwidth} to \ref{SBS_computation}, the proposed algorithm, which considers latency, energy consumption and learning performance, and can jointly optimize multiple wireless resources, can always achieve better performance.
Moreover, it is robust to the change of these system parameters.

In different network environments, the system parameter $\rho$ determined by all federal learning participants may be different. 
By analyzing the influence of the weighted parameter in figure \ref{weight}, we can find that the difference of the weighted parameter has an effect on the performance of each algorithm. 
However, it is worth noting that the proposed algorithm can always provide performance gains for different a values compared with the benchmarks, which further confirms the performance of the proposed algorithm..

\section{Conclusion And Future Works}
In this paper, we investigate federated learning scenarios in hierarchical networks.
The minimization of the weighted sum of the system cost and learning cost is programmed as a MINLP problem.
Then the joint optimization problem is decomposed into several sub-problems, and an effective algorithm is designed for each sub-problem.
Therefore the original problem is solved by iteratively solving each subproblem.
Finally, the numerical results show that the proposed algorithm can quickly converge to a stable solution and achieve better performance compared with the benchmarks under various parameter settings.

Asynchronous federated learning tends to have better performance over parameter aggregation latency than synchronous federated learning considered in this paper.
Therefore, it is promising to study the resource allocation algorithm under the asynchronous federated learning scenario based on the algorithm designed in this paper.
Meanwhile, the impact of different data distribution among federated learning organizations on resource allocation and the avoidance of toxic data affecting learning performance through intelligent data selection are also worthy of further discussion and research.

\appendix 
\subsection{Proof of Proposition 1}
By substituting formula (\ref{rate_appendix_a}) into (\ref{time_appendix_a}), we can get
\begin{equation}\label{appendix_a_1}
t_{k}=\frac{D_{k}^{\rm s}}{B_{k}^{\rm s}{\rm log}_{2}\left(1+\frac{p^{\rm max}h_k^{\rm s}}{B{_{j}}N_0}\right)}.
\end{equation}
The optimization variable contained in the above equation is $B_k$, so (\ref{appendix_a_1}) can be abbreviated as
\begin{equation}\label{appendix_a_2}
t_{k}=\frac{\hat{D}_{k}^{\rm s}}{B_k^{\rm s}},
\end{equation}
where $\hat{D}_{k}^{\rm s}=\frac{D_{k}^{\rm s}}{log_{2}\left(1+\frac{p^{\rm max}h_k}{B{_{j}}N_0}\right)}$ is a constant independent of the optimization variable.

Next we will start the process of proof by contradiction.
First we assume that for the SBS $j$, which satisfies $t_j^{\rm total}=t^{*}$, the time it takes to receive data from all sensors is not equal.
Note that $t^*$ denotes the shortest time to complete a round of federated learning.
Without loss of generality, we consider the case where there are two sensors(sensor 1 and sensor 2) within its range.
Further, we assume that the time to receive the data of sensor 1 is longer than that of sensor 2, which can be expressed as
\begin{equation}\label{appendix_a_3}
\frac{\hat{D}_{1}^{\rm s}}{B_1^{\rm s}} \textgreater \frac{\hat{D}_{2}^{\rm s}}{B_2^{\rm s}}.
\end{equation} 
Now the delay of the data receiving process $t_j^{\rm r}=\frac{\hat{D}_{1}^{\rm s}}{B_1^{\rm s}}$.
Besides, in order to minimize the delay of the data receiving process, it is easy to obtain the following equation
\begin{equation}\label{appendix_a_4}
B_1^{\rm s}+B_2^{\rm s}=B_j.
\end{equation} 

So if we put (\ref{appendix_a_3}) and (\ref{appendix_a_4}) together, we get that if we increase $B_1^{\rm s}$ and decrease $B_2^{\rm s}$ appropriately, $t_1$ and $t_2$ are going to go down to $\hat{t}_1$ and up to $\hat{t}_2$ respectively.
If we only make minor adjustments to the allocation of bandwidth resources, so $\hat{t}_1 \textgreater \hat{t}_2$ still holds.
However, now $\hat{t}_j^{\rm r}=\hat{t}_1 \textless t_1$ is satisfied.
So when $t_j^{\rm cmp}$ and $t_j^{\rm up}$ stay the same, the formula $\hat{t}^{\rm one}=\hat{t}_j^{\rm r}+t_j^{\rm cmp}+t_j^{\rm up} \textless t^{*}$ is got.
This contradicts the false fact that $t^{*}$ is the optimal time, so the original assumption is invalid.
Besides the results can be easily extended to scenarios where multiple sensors exists.

That's the end of the proof.

\subsection{Proof of Theorem 1}
We first solve for $B_k^{\rm s*}$ in terms of $a_j^*$ by the transformation of the equation:

\begin{equation}
   \frac{D_{k}^{\rm s}}{{B_{k}^{\rm s}}^*{\rm log}_{2}\left(1+p^{\rm max}h_{k}^{\rm s}/B{_{j}}N_{0}\right)}=a{_{j}^{*}}
\end{equation}
$\Rightarrow$
\begin{equation}\label{a_present_b}
	B_{k}^{\rm s*}=\frac{D_k^{\rm s}}{a_j^{*}{\rm log}_2\left(1+p^{\rm max}h_k^{\rm s}/B_j{N_0}\right)}.
\end{equation}

We can substitute this expression for b into another equation, thus we can get
\begin{equation}
    \frac{1}{a_j^*}\left(\sum_{{i}\in\mathcal{S}_{j}}D_{i}^{\rm s}/{\rm log}_{2}\left(1+p^{\rm max}h_{i}^{\rm s}/B_j{N_0}\right)\right)=B_j.
\end{equation}
It's easy to figure it out from the above equation
\begin{equation}
	a{_{j}^{*}}=\frac{\sum_{{i}\in\mathcal{S}_{j}}D_{i}^{\rm s}/{\rm log}_{2}\left(1+\frac{p^{\rm max}h_{i}^{\rm s}}{B{_{j}}N_{0}}\right)}{B_j}.
\end{equation}
Bring a back into equation (\ref{a_present_b}), we have
\begin{equation}
		{B_{k}^{\rm{s}}}^*=\frac{B_jD_{k}^{\rm s}/{\rm log}_{2}\left(1+\frac{p^{\rm max}h_{k}^{\rm s}}{B{_{j}}N_{0}}\right)}{\sum_{{i}\in\mathcal{S}_{j}}D_{i}^{\rm s}/{\rm log}_{2}\left(1+\frac{p^{\rm max}h_{i}^{\rm s}}{B{_{j}}N_{0}}\right)},\ \forall{k}\in\mathcal{S}_{j}.
\end{equation}

That's the end of the proof.

\bibliographystyle{IEEEtran}
\bibliography{IEEEabrv,ref}

\end{document}